\documentclass[acmsmall]{acmart}
\usepackage{multirow}
\AtBeginDocument{%
  \providecommand\BibTeX{{%
    \normalfont B\kern-0.5em{\scshape i\kern-0.25em b}\kern-0.8em\TeX}}}
\usepackage{graphicx}
\setcopyright{acmcopyright}
\copyrightyear{2021}
\acmYear{2021}
\acmDOI{10.1145/1122445.1122456}
\usepackage{cleveref}
\usepackage{amsmath,subcaption,multirow}
\usepackage{comment}
\usepackage{wrapfig}
\newtheorem*{dfn*}{Problem Formulation}

\DeclareMathOperator*{\argmax}{arg\,max}
\DeclareMathOperator*{\argmin}{arg\,min}

\usepackage[ruled,linesnumbered]{algorithm2e}
\newcommand{\nonl}{\renewcommand{\nl}{\let\nl\oldnl}}
\makeatletter
\newcommand{\nosemic}{\renewcommand{\@endalgocfline}{\relax}}
\acmJournal{JACM}
\acmVolume{37}
\acmNumber{4}
\acmArticle{111}
\acmMonth{9}

\begin{document}

% %% allowing the author to define a "short title" to be used in page headers.
% \title{Deep Reinforcement Learning in Dynamic Recommender Systems: Challenges, Recent Advances and Future Directions}

% \title{Deep Reinforcement Learning in Recommender Systems: Challenges, Recent Advances and Future Directions}

\title{A Survey of Deep Reinforcement Learning in Recommender Systems: A Systematic Review and Future Directions}


\author{Xiaocong Chen}
\email{xiaocong.chen@unsw.edu.au}
\affiliation{%
  \institution{University of New South Wales}
  \city{Sydney}
  \state{NSW}
  \country{Australia}
}

\author{Lina Yao}
\email{lina.yao@unsw.edu.au}
\affiliation{%
  \institution{University of New South Wales}
  \city{Sydney}
  \state{NSW}
  \country{Australia}
  }

\author{Julian Mcauley}
\email{jmcauley@eng.ucsd.edu}
\affiliation{%
  \institution{University of California, San Diego}
  \state{CA}
  \country{USA}
}

\author{Guanglin Zhou}
\email{guanglin.zhou@unsw.edu.au}
\affiliation{%
  \institution{University of New South Wales}
  \city{Sydney}
    \state{NSW}
  \country{Australia}
}

\author{Xianzhi Wang}
\email{xianzhi.wang@uts.edu.au}
\affiliation{%
  \institution{University of Technology Sydney}
  \city{Sydney}
    \state{NSW}
  \country{Australia}
}

%\renewcommand{\shortauthors}{Chen et al.}

\begin{abstract}
%   %With the growing volume of online information, recommender systems receive plenty of research interest in the past decade. 
%   Recommender systems are behind various web applications to help find information or items that fit users' interest. Recommender systems can be considered as an interactive process: the system recommends items to users, users provide feedback, and the system updates itself based on users' feedback. An important thread of research formulates the above as a Markov Decision Process and employs reinforcement learning to conduct recommendations. %Deep reinforcement learning has received growinghttps://www.overleaf.com/project/600f7452f541326618642de8 interest in recommender systems research in recent years.
%   Deep reinforcement learning combines the advantages of reinforcement learning and deep learning,
%   %achieves superiority progress in recommender systems. 
%   leading to significant progress in recommender systems.
%   In light of the emergence of deep reinforcement learning (DRL) in recommendation research and the fruitful results in recent years, this article aims to provide a timely and comprehensive review of recent advances in DRL-based recommender systems.
%   We start with the motivation of applying DRL in recommender systems.
%   Then, we provide a taxonomy of DRL-based recommender systems and a summary of existing methods. We highlight the emerging topics in this area of research, summarize open issues, and comment on future directions for DRL-based recommender systems.

In light of the emergence of deep reinforcement learning (DRL) in recommender systems research and several fruitful results in recent years, this survey aims to provide a timely and comprehensive overview of the recent trends of deep reinforcement learning in recommender systems. We start with the motivation of applying DRL in recommender systems. Then, we provide a taxonomy of current DRL-based recommender systems and a summary of existing methods. We discuss emerging topics and open issues, and provide our perspective on advancing the domain. This survey serves as introductory material for readers from academia and industry into the topic and identifies notable opportunities for further research.

\end{abstract}



\maketitle
% \newpage
% \setcounter{tocdepth}{3}
% \tableofcontents
\section{Introduction}
% Recommender Systems are a type of information system which aims to learn users' preferences from historical feedback. 
%With the exponential growing online information, users are facing numerous items, movies or products etc. 
%As a result, the 
% Recommender System aims to learn users' preferences from historical feedback. 
% Personalized strategies are essential to improve user experience when using web applications such as e-commerce. 
% %In a word, the 
% Recommender system are a crucial component for existing commercial applications and play an important role in improving users' experience. 
%Deep learning is widly used in 

%Recommender systems have been a longstanding topic of sustained interest to the research community and are important to a wide range of domains and practical applications.
Recent years have seen significant progress in recommendation techniques, from traditional recommendation techniques, e.g., collaborative filtering, content-based recommendation and matrix factorization~\cite{lu2015recommender}, to deep learning based techniques.
In particular, deep learning show strong advantages in solving complex tasks and dealing with complex data, due to its capability to capture non-linear user-item relationships and deal with various types of data sources such as images and text.
It has thus been increasingly used in recommender systems.
%which 
Deep learning-based recommender systems have limitations in capturing interest dynamics~\cite{chen2020knowledge,zhang2019deep} due to distribution shift, i.e., the training phase is based on an existing dataset which may not reflect real user preferences that undergo rapid change.
In contrast, deep reinforcement learning (DRL) aims to train an agent that can learn from interaction trajectories provided by the environment by combining the power of deep learning and reinforcement learning. Since an agent in DRL can actively learn from users' real-time feedback to infer dynamic user preferences,
%of users, 
DRL is especially suitable for learning from interactions, such as human-robot collaboration; it has also driven significant advances in a range of interactive applications ranging from video games, Alpha Go to autonomous driving~\cite{arulkumaran2017deep}.
%DRL is designed 
%DRL, that integrates deep learning and reinforcement learning, now receives growing interest to be applied in recommender systems. 
%DRL can also mitigate data sparsity issues and improve recommendation performance.
In light of the significance and recent progresses in DRL for recommender sytsems, we aim to timely summaize and comment on DRL-based recommendation systems in this survey.
%is gaining increasing attention in enhancing recommender systems. 
%Distribution shift is caused by the distribution discrepancy between the training set and test set but they are not available in DRL.

%\noindent\textbf{What are the differences between this survey and existing ones?}
% There are a few existing surveys which mention DRL-based recommender systems~\cite{afsar2021reinforcement}. However, 
% There are few comprehensive reviews which categorize and position current progress and shed a light on future direction in this fast growing area~\cite{afsar2021reinforcement}. 

%%%% TO DO
%The 
A
recent survey on reinforcement learning based recommender systems \cite{afsar2021reinforcement} provides a general review about reinforcement learning in recommender systems without a sophsiticated investigation of the growing area of deep reinforcement learning. Our survey distinguishes itself in providing a systematic and comprehensive overview of existing methods in DRL-based recommender systems, along with a discussion of emerging topics, open issues, and future directions. This survey introduces researchers, practitioners and educators into this topic
%, and build a blueprint of deep reinforcement learning in recommender systems 
and fostering an understanding of the key techniques in the area.
The main contributions of this survey include the following:
\begin{itemize}
    \item We provide an up-to-date  comprehensive review of deep reinforcement learning in recommender systems, with state of the art techniques and pointers to core references. To the best of our knowledge, this the first comprehensive survey in deep reinforcement learning based recommender systems.
    \item We present a taxonomy of the literature of deep reinforcement learning in recommender systems. Along with the outlined taxonomy and literature overview, we discuss the benefits, drawbacks and give suggestions for future research directions. 
    \item We shed light on emerging topics and open issues for DRL-based recommender systems. We also point out future directions that could be crucial for advancing DRL-based recommender systems.
\end{itemize}

The remainder of this survey is organized as follows: Section 2 provides an overview of recommender systems, DRL and their integration. Section 3 provides a literature review with a taxonomy and classification mechanism. Section 4 reviews emerging topics, and Section 5 points out open questions. Finally, Section 6 provides a few promising future directions for further advances in this domain.

\begin{figure}
    \centering
    \includegraphics[width=\linewidth]{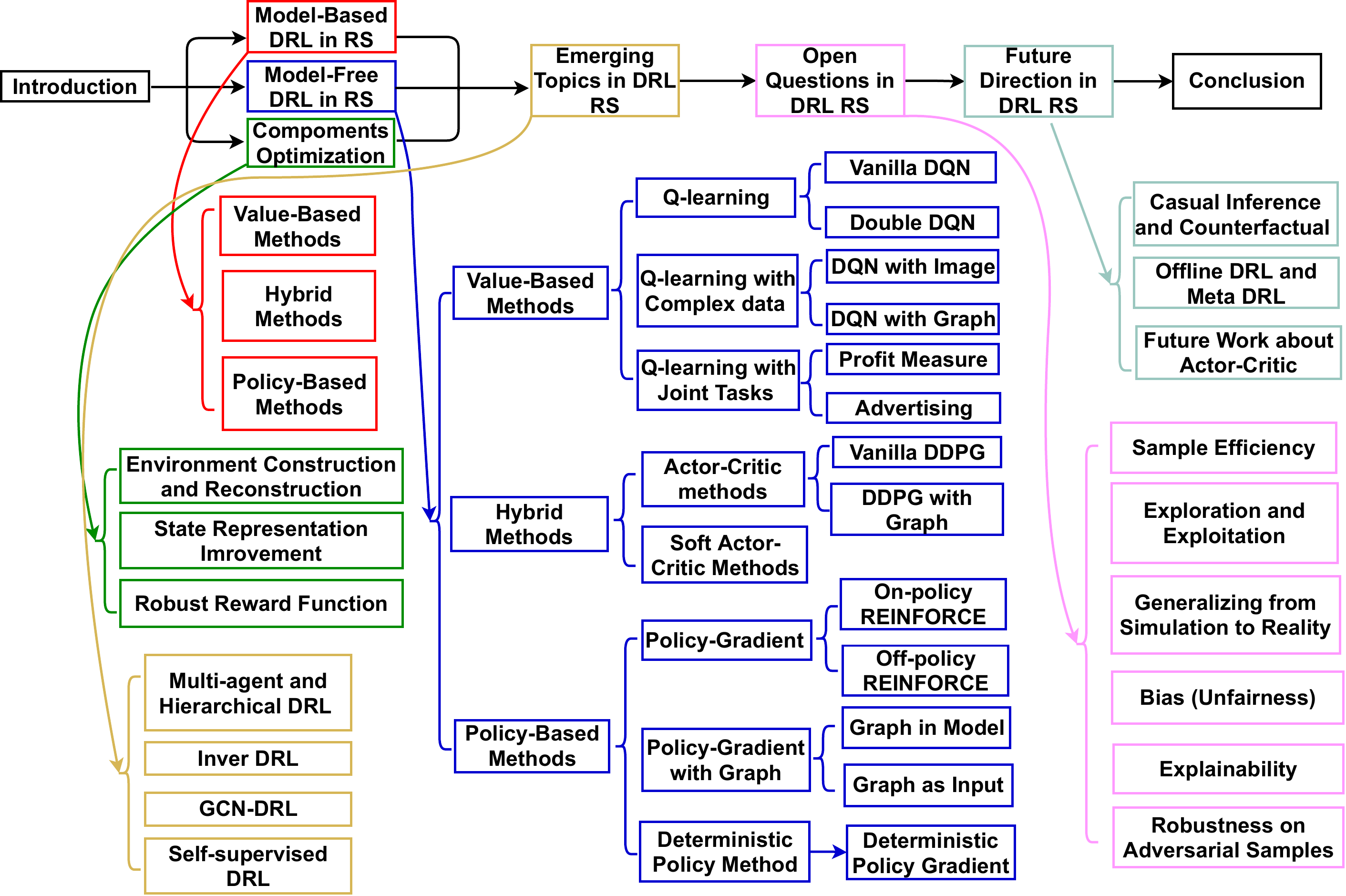}
    \caption{Taxonomy of Deep Reinforcement Learning based Recommender Systems in this survey}
    \label{fig:overview}
\end{figure}
%\section{Overview of Dynamic Recommender Systems and DRL}

%\subsection{Dynamic Recommender Systems}

\section{Background}

%Before we dive into the details of this survey, 
In this section,we introduce key concepts related to dynamic recommender systems (RS) and deep reinforcement learning (DRL), and motivate the introduction of DRL to dynamic recommender systems.
%and provide a problem statement for DRL-based RS.

\subsection{Why Deep Reinforcement Learning for Recommendation?}

% \textcolor{red}{we change a logic in this section}

%%%%%%%%%%%%%%%%

%%% don't need to emphasize dynamic here, just say DRL is able to capture the dynamics of human perferences and interest something like this
Recommender systems require coping with \emph{dynamic} environments by estimating rapidly changing users' preferences and proactively recommending items to users.
%Moreover, Dynamic recommender systems require to update themselves periodically based on users' feedback. 
Let $\mathcal{U}$ be a set of users of cardinality $|\mathcal{U}|$ and $\mathcal{I}$ be a set of items of cardinality $|\mathcal{I}|$.
%The sets of users and items are assumed to be fixed. 
For each user $u\in\mathcal{U}$, we observe a sequence of user actions $\mathbb{X}^u = [x_1^u, x_2^u, \cdots , x_{T_u}^u]$ with item $x_t^u \in \mathcal{I}$, i.e., each event in a user sequence comes from the item set. We refer to a user making a decision as an interaction with an item. Suppose the feedback (e.g., ratings or clicking behavior) provided by users is $\mathcal{F}$, then a dynamic recommender system maintains the corresponding recommendation policy $\pi^u_t$, which will be updated systematically based on the feedback $f^u_i \in \mathcal{F}$ received during the interaction for item $i \in \mathcal{I}$ at the timestamp $t$. 

\begin{figure}[ht]
    \centering
    \begin{subfigure}[b]{0.68\textwidth}
         \centering
         \includegraphics[width=\textwidth]{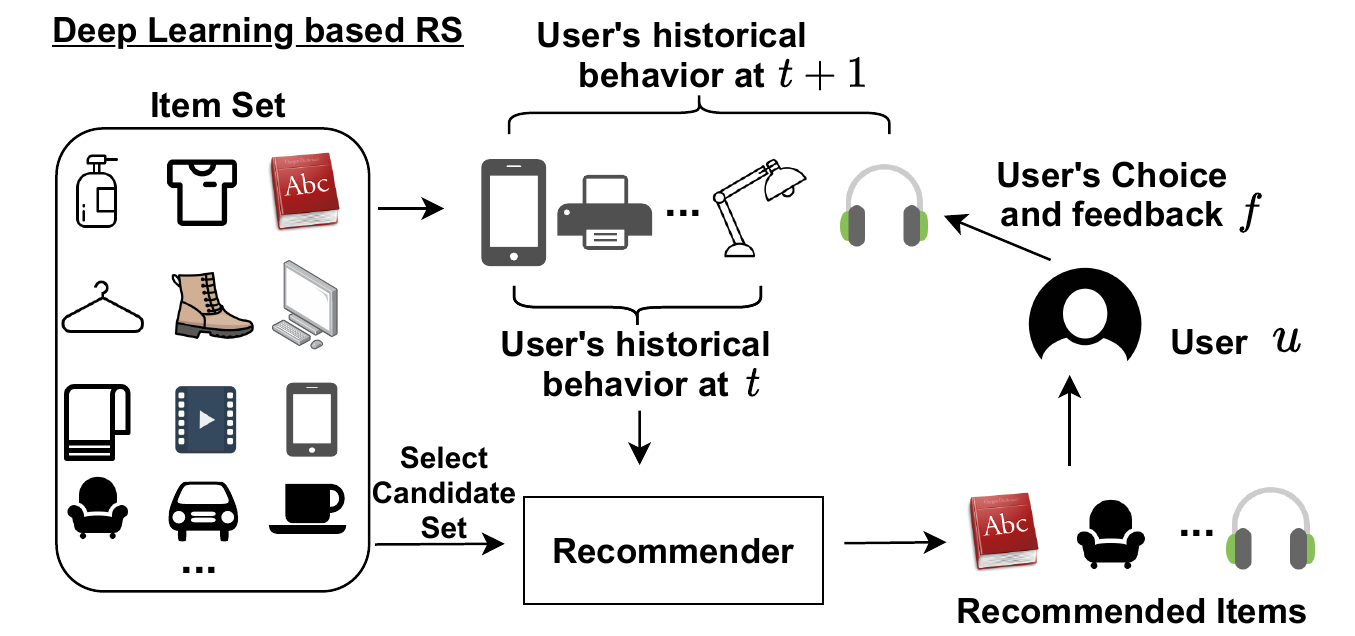}
         \caption{Deep learning based recommender systems}
     \end{subfigure}
         \begin{subfigure}[b]{0.65\textwidth}
         \centering
         \includegraphics[width=\textwidth]{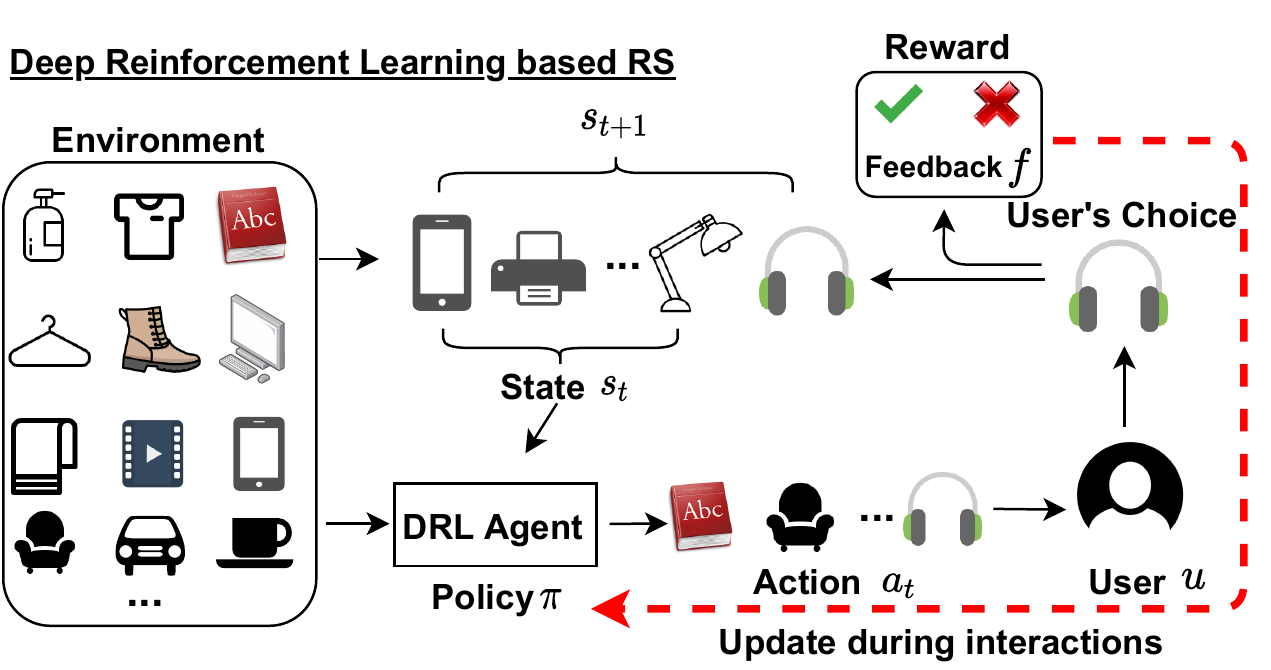}
         \caption{Deep reinforcement learning based recommender systems}
     \end{subfigure}
    \caption{Difference between deep learning based RS and DRL-based RS. Deep learning based RSs may only update the recommendation policy during the training stage. They often require re-training, which is computationally inefficient, when users' interests change significantly. DRL-based RS will update the recommendation policy time over time as new rewards are received.}
    \label{fig:catnew}
\end{figure}

The marriage of deep learning and reinforcement learning has fueled breakthroughs in recommender systems. 
% The major process can be pipelined in three stages includes user data collection including historical data and demographic data, user/item embedding and recommender. Similarly, DRL-based dynamic RS can be concluded into three stages as well. 
DRL-based RS consists of a pipeline with three building blocks: environment construction, state representation and recommendation policy learning. Environment construction aims to build an environment based on a set of users' historical behaviors. State representation is provided by the environment containing certain user information including historical behavior, demographic data (etc.). Recommendation policy learning is the key component to understand and predict users' future behavior. 
DL-based RS receives user feedback (e.g., ratings or clicks) to reflect users' interests and update the recommender, while DRL-based RS receives the reward provided by the environment to update the policy. The reward provided by the environment is a pre-defined function containing several factors. The detailed process of DL based RS and DRL-based RS mapping can be found in~\Cref{fig:catnew}.

\subsection{Preliminaries of Deep Reinforcement Learning}
% Deep reinforcement learning is considered to be the combination of deep learning and reinforcement learning. 
The typical defining feature of DRL is to use the deep learning to approximate reinforcement learning's value function and solve high-dimensional Markov Decision Processes (MDPs)~\cite{arulkumaran2017deep}. Formally, a MDP can be represented as a tuple ($\mathcal{S},\mathcal{A},\mathcal{P},\mathcal{R},\gamma$). The agent chooses an action $a_t\in\mathcal{A}$ according to the policy $\pi_t(s_t)$ at state $s_t\in\mathcal{S}$. The environment receives the action and produces a reward $r_{t+1}\in\mathcal{R}$ and transfers the reward into the next state $s_{t+1}$ according to the transition probability $P(s_{t+1}|s_t,a_t)\in\mathcal{P}$. The transition probability $\mathcal{P}$ is unknown beforehand in DRL.
%domain. 
Such a process continues until the agent reaches the terminal state or exceeds a pre-defined maximum time step. The overall objective is to maximize the expected discounted cumulative reward,
\begin{align}
    \mathbb{E}_\pi [r_t] = \mathbb{E}_\pi \big[\sum_{0}^\infty \gamma^k r_{t+k}\big] 
\end{align}
where $\gamma \in [0,1]$ is the discount factor that balances the future reward and the immediate reward.
\begin{wrapfigure}{r}{0.45\textwidth}
      \begin{center}
        \includegraphics[width=0.45\textwidth]{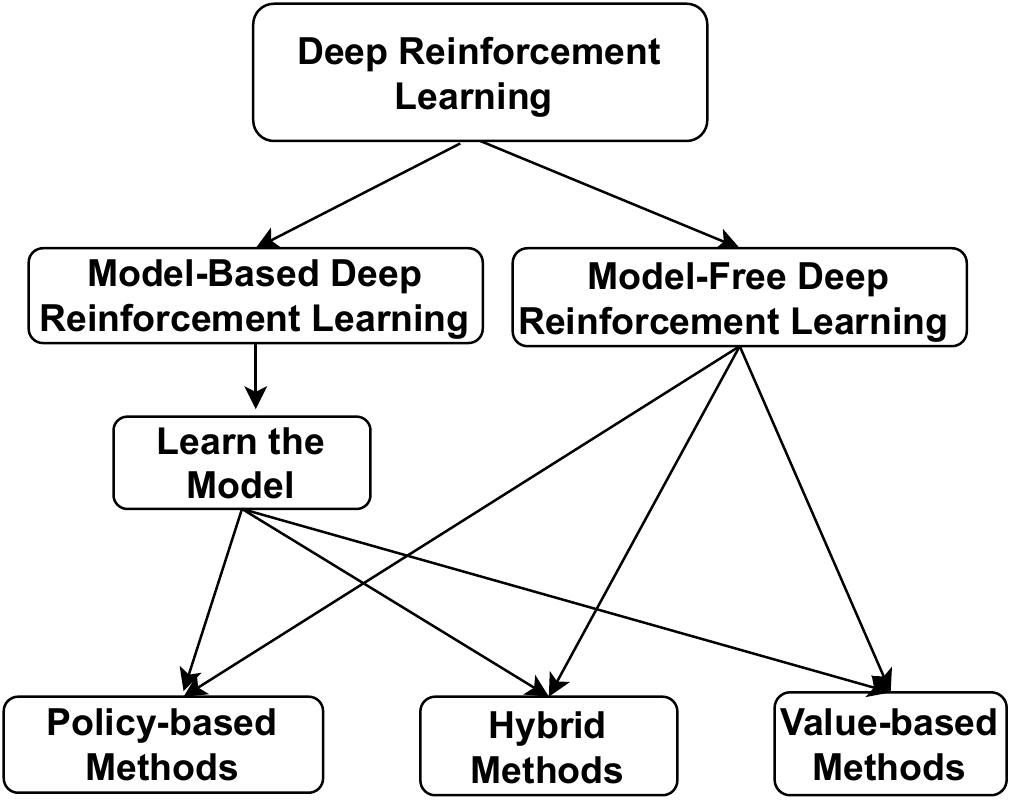}
      \end{center}
    \caption{Taxonomy of Deep Reinforcement Learning in Recommender Systems}
    \label{fig:tax}
\end{wrapfigure}

Deep reinforcement learning can be divided into two categories: \emph{model-based} and \emph{model-free} methods (a detailed taxonomy can be found in \Cref{fig:tax}). The major {\emph difference} between the two is whether the agent can learn a model of the environment. Model-based methods aim to estimate the transition function and reward function, while model-free methods aim to estimate the value function or policy from experience. In model-based methods, the agent accesses the environment and plans ahead while model-free methods gain sample efficiency from using models which are more extensively developed and tested than model-based methods in recent literature~\cite{arulkumaran2017deep}. 

Deep reinforcement learning approaches are divided into three streams: \emph{value-based}, \emph{policy-based} and \emph{hybrid} methods. In value-based methods, the agent updates the value function to learn a policy; policy-based methods learn the policy directly; and hybrid methods combine value-based and policy-based methods called \emph{actor-critic} methods. Actor-critic contains two different networks where an actor network uses a policy-based method and the critic uses a value-based method to evaluate the policy learned by the agent.
\begin{table}[ht]
    \centering
    \caption{Notations}
    \begin{tabular}{cc|ccc}
        \hline
        Notations & Name & Notations & Name & Notes \\\hline
         $Q(\cdot)$  &  Q-Value Function   &           $s$ &   State     &  users' preference     \\
         $V(\cdot)$  &  Value Function    &           $a$ &   Action      &  Recommended item(s)    \\
         $\gamma$ & Discount Factor      &           $\pi$, $\mu(\cdot)$  & Policy  & Recommendation policy       \\
          $\mathbb{E}$ & Expected Value      &          $r(\cdot,\cdot)$ & Reward   &  users' click behavior    \\
          $\theta$ & Model Parameter      &           $\alpha$ & constant $\in [0,1]$      &   -    \\
          $p(\cdot)$ & Transition Probability    &  $\tau$ & Sampled Trajectory      &  A tuple $(s_t,a_t,s'_t,r_t)$    \\\hline
    \end{tabular}
    \label{tab:my_label}
\end{table}

Deep reinforcement learning can be divided into \emph{on-policy} and \emph{off-policy} methods.
In off-policy methods,
%mean that the behavior policy used for selecting actions is different from the target policy. Normally, 
the behavior policy $\pi_b$ is used for exploration while the target policy $\pi$ is used for decision-making.
For on-policy methods, the behavior policy is the same as the target policy. 

\textbf{Q-learning}~\cite{watkins1992q} is an off-policy value-based learning scheme for finding a greedy target policy:
\begin{align}
    \pi(s) = \argmax_a Q_\pi (s,a)
\end{align}
where $Q_u (s,a)$ denotes the $Q$-value and is used in a small discrete action space. For a deterministic policy, the $Q$ value can be calculated as follows
\begin{align}
    Q (s_t,a_t) = \mathbb{E}_{\tau \sim \pi}[r(s_t,a_t) + \gamma Q(s'_t, a'_t)].
\end{align}
Deep Q learning (DQN)~\cite{mnih2015human} uses deep learning to approximate a non-liner Q function parameterized by $\theta_q$: $Q_{\theta_q} (s,a)$. DQN designs a network $Q_{\theta_q}$ that is asynchronously updated by minimizing the MSE:
\begin{align}
    \mathcal{L}(\theta_q) = \mathbb{E}_{\tau \sim \pi}\Big[Q_{\theta_q}(s_t,a_t)-(r(s_t,a_t) + \gamma Q_{\theta_q}(s'_{t},a'_{t}))\Big]^2 \label{dqnloss}
\end{align}
where $\tau$ is the sampled trajectory containing $(s,a,s',r(s,a))$. In particular, $s'_t$ and $a'_t$ come from the behavior policy $\pi_b$ while $s,a$ comes from the target policy $\pi$.
%Moreover, 
It is worth mentioning that the value function $V_\pi(s)$ represents the expected return. $V_\pi(s)$ is used to evaluate the goodness of the state while $Q_\pi(s_t,a_t)$ is used to evaluate the action. $V_\pi(s)$ can be defined as
\begin{align}
    V_\pi(s) = \mathbb{E}_{\tau\sim \pi}\bigg[\sum_{t=0}^T\gamma^t r(s,a)|s_0=s\bigg].
\end{align}

$V_\pi(\cdot)$ and $Q_\pi(\cdot)$ have the following relationship:
%which are widely used in value based works:
\begin{align}
    V_\pi(s) = \mathbb{E}_{a\sim\pi}[Q_\pi(s,a)].
\end{align}

The value function is updated using the following rule with the Temporal Difference (TD) method,
\begin{align}
    V_\pi(s_t) \leftarrow V_\pi(s_t) + \alpha[\underbrace{r(s'_t,a'_t) + \gamma V_\pi(s'_t) - V_\pi(s_t)}_{\text{TD-error}}]
\end{align}
where $\alpha$ is a constant.

\textbf{Policy gradient}~\cite{williams1992simple} is an on-policy policy-based method which can handle high-dimensional or continuous actions which cannot be easily handled by Q-learning. Policy gradient
%is the most popular policy-based method that
aims to find the parameter $\theta$ of $\pi_{\theta}$ to maximize the accumulated reward. To this end, it maximizes the expected return from the start state:
\begin{align}
    J(\pi_\theta) = \mathbb{E}_{\tau \sim \pi_{\theta}}[r(\tau)] = \int\pi_{\theta}(\tau) r(\tau)d\tau
\end{align}
where $\pi_{\theta}(\tau)$ is the probability of the occurrence of $\tau$. Policy gradient learns the parameter $\theta$ by the gradient $\nabla_\theta J(\pi_\theta)$ as defined below:
\begin{align}
    \nabla_\theta J(\pi_\theta) = \int\pi_{\theta}(\tau) r(\tau)d\tau &=  \int\pi_{\theta}(\tau) \nabla_{\theta}\log \pi_{\theta}(\tau)r(\tau)d\tau \notag \\
    & =\mathbb{E}_{\tau \sim d_{\pi_\theta}}[\sum_{t=1}^Tr(s_t,a_t)\sum_{t=1}^T\nabla_{\theta} \log \pi_{\theta}(s_t,a_t)].
\end{align}
The above derivations contain the following substitution,
\begin{align}
    \pi_{\theta}(\tau) = p(s_1)\prod_{t=1}^T \pi_{\theta}(s_t,a_t)p(s_{t+1}|s_t,a_t)
\end{align}
where $p(\cdot)$ are independent from the policy parameter $\theta$, which is omitted during the derivation.
%It is worth mentioning that the famous policy gradient algorithm REINFORCE uses 
Monte-Carlo sampling has been used by previous policy gradient algorithm (e.g,. REINFORCE) for $\tau\sim d_{\pi_{\theta}}$.

\textbf{Actor-critic networks} combine the advantages from Q-learning and policy gradient. They can be either on-policy~\cite{konda2000actor} or off-policy~\cite{degris2012off}. An actor-critic network consists of two components: i) \textit{an actor}, which optimizes the policy $\pi_\theta$ under the guidance of $\nabla_\theta J(\pi_{\theta})$; and ii) \textit{a critic}, which evaluates the learned policy $\pi_\theta$ by using $Q_{\theta_q} (s,a)$. The overall gradient is represented as follows:
\begin{align}
    \mathbb{E}_{s \sim d_{\pi_\theta}}[Q_{\theta_q}(s,a)\nabla_{\theta} \log \pi_{\theta}(s,a)].
\end{align}

When dealing with off-policy learning, the value function for $\pi_\theta (a|s)$ can be further determined by deterministic policy gradient (DPG) as shown below:
\begin{align}
    \mathbb{E}_{s \sim d_{\pi_\theta}}[\nabla_a Q_{\theta_q}(s,a)|_{a=\pi_\theta(s)}\nabla_{\theta} \pi_{\theta}(s,a)].
\end{align}

While traditional policy gradient calculates the integral for both the state space $\mathcal{S}$ and the action space $\mathcal{A}$, DPG only requires computing the integral to the state space $\mathcal{S}$.
Given a state $s\in\mathcal{S}$, there will be only one corresponding action $a\in\mathcal{A}:\mu_\theta(s) = a$ using DPG. Specifically, deep Deterministic Policy Gradients (DDPG) is an algorithm that combines techniques from DQN and DPG. DDPG contains four different neural networks: Q Network $Q$, policy network, target Q network $Q^{\mathit{tar}}$, and target policy network. It uses the target network for both the Q Network $Q$ and policy network $\mu$ to ensure stability during training. Assume $\theta_q, \theta_{\pi}, \theta_{q'}$ and $\theta_{\pi'}$ are parameters of the above networks; then DDPG soft-updates the parameters for the target network~\cite{lillicrap2015continuous}:
\begin{align}
    \text{Actor: }\theta_{\pi'} \leftarrow \alpha \theta_\pi + (1-\alpha)\theta_{\pi'}  \text{ Critic: }\theta_{q'} \leftarrow \alpha \theta_q + (1-\alpha)\theta_{q'} 
\end{align}
%Notably, in some literature, actor-critic is categorized as policy-base methods as the actor is the decision-maker. In this survey, we categorize it as a hybrid method because of its structure.

\subsection{DRL meets RS: Problem Formulation}
DRL is normally formulated as a Markov Decision Process (MDP).
Given a set of users $\mathcal{U} = \{u, u_1, u_2, u_3, ...\}$, a set of items $\mathcal{I} = \{i, i_1, i_2, i_3, ...\}$, the system first recommends item $i$ to user $u$ and then gets feedback $f_i^u$.
The system aims to incorporate the feedback to improve future recommendations and needs to determine an optimal policy $\pi^*$ regarding which item to recommend to the user to achieve positive feedback.
The MDP modelling of the problem treats the user as the environment and the system as the agent. The key components of the MDP in DRL-based RS include the following:
\begin{itemize}
    \item State $\mathcal{S}$: A state $S_t\in\mathcal{S}$ is determined by both users' information and the recent $l$ items in which the user was interested before time $t$.
    %Demographic information may also be included.
    \item Action $\mathcal{A}$: An action $a_t \in\mathcal{A}$ represents users' dynamic preference at time $t$ as predicted by the agent. $\mathcal{A}$ represents the whole set of (potentially millions of) candidate items.
    \item Transition Probability $\mathcal{P}$: The transition probability $p(s_{t+1}|s_t,a_t)$ is defined as the probability of state transition from $s_t$ to $s_{t+1}$ when action $a_t$ is executed by the recommendation agent. In a recommender system, the transition probability refers to users' behavior probability. $\mathcal{P}$ is only used in model-based methods.
    \item Reward $\mathcal{R}$: Once the agent chooses a suitable action $a_t$ based on the current state $S_t$ at time $t$, the user will receive the item recommended by the agent. Users' feedback on the recommended item accounts for the reward $r(S_t,a_t)$. The feedback is used to improve the policy $\pi$ learned by the recommendation agent.
    \item Discount Factor $\gamma$: The discount factor $\gamma \in [0,1]$ is used to balance between future and immediate rewards---the agent focuses only on the immediate reward when $\gamma=0$ and takes into account all the (immediate and future) rewards otherwise.
\end{itemize}
%With the aforementioned definitions, 

The DRL-based recommendation problem can be defined by using MDP as follows.
\textit{Given the historical MDP, i.e., $(\mathcal{S},\mathcal{A},\mathcal{P}, \mathcal{R},\gamma)$, the goal is to find a set of recommendation polices ($\{\pi\}: \mathcal{S}\to \mathcal{A}$) that maximizes the cumulative reward during interaction with users.}
%In DRL-based RS, such process can be described as:
\begin{dfn*}
Given an environment that contains all items $\mathcal{I}$, when user $u\in\mathcal{U}$ interacts with the system, an initial state $s$ is sampled from the environment which contains a list of candidate items and users' historical data. The DRL agent needs to work out a recommendation policy $\pi$ based on the state $s$ and produces the corresponding recommended item list $a$.
The user will provide feedback on the list which is normally represented as click or not click. The DRL agent will then utilize the feedback to improve the recommendation policy and move to the next interaction episode.
\end{dfn*}

\section{Deep Reinforcement Learning in Recommender Systems}

DRL-based RS has some unique challenges such as state construction, reward estimation and environment simulation. We categorize the existing work of DRL-based recommendation into model-based and model-free methods (the taxonomy is shown in~\Cref{fig:tax}).

% We classify existing methods which employs DRL in RS into two different streams: holistic recommendation~\Cref{sec:holistic} and component optimization~\Cref{sec:component}. Holistic recommedation 
% similar to traditional DRL, 

\subsection{Model-based Deep Reinforcement Learning based Methods}
%Model-based methods require the agent to access the environment and plan ahead which is challenging in recommendation scenarios.
Model-based methods assume an expected reward or action available for the next step to help the agent update the policy.
%To ensure consistency, we generally categorize them into three folds as well: policy-based, value-based, and hybrid methods. 
\begin{table}[ht]
\caption{List of publications in model-based DRL-based RS}
\begin{tabular}{ccc}
\hline
Method  & Work \\\hline
Value-based &  \cite{zhang2017dynamic,chen2019generative,zou2020pseudo,wang2021reinforcement} \\
Policy-based &  \cite{bai2019model,hong2020nonintrusive}\\
Hybrid &  \cite{zhao2020whole}
\\\hline
\end{tabular}
\end{table}

\paragraph{Policy-based methods}
IRecGAN~\cite{bai2019model} is a model-based method that adopts generative adversarial training to improve the robustness of policy learning. It can reduce the cost of interaction for RS by using offline data instead of the simulated environment.
%Benefit from the offline setting, the expect reward and action come to reachable. 
IRecGAN employs a generative adversarial network~\cite{goodfellow2014generative} to generate user data based on the offline dataset.
It trains a recommendation agent using a policy gradient-based DRL method called REINFORCE. The agent aims to learn a policy based on the following gradient,
\begin{align}
    \mathbb{E}_{\tau\sim\{g,data\}}\big[\sum_{t=0}^T \sum_{t'=t}^T\gamma^{t'-t} q_D(\tau_{0:t}^n)r_t  \nabla_{\theta_a}(c_t \in \pi_{\theta_a}(s_t)\big],q_D(\tau_{0:t}^n)= \frac{1}{N}\sum_{n=1}^N D(\tau_{0:T}^n),\tau_{0:T}^n\in MC^{\mathcal{U}}(N)
\end{align}
where the $\mathit{MC}^{\mathcal{U}}(N)$ represents the sampled $N$ sequences from the interaction between $\mathcal{U}$ and the agent using the Monte-Carlo tree search algorithm, $D$ is the discriminator, $T$ is the length of $\tau$, $g$ represents the offline data, and \textit{data} represents the generated data. 

\citet{hong2020nonintrusive} propose NRSS for personalized music recommendation. NRSS uses wireless sensing data to learn users' current preferences. NRSS considers three different types of feedback: score, option, and wireless sensing data. Because multiple factors are considered as the reward, NRSS designs a reward model which consists of users' preference reward $r_p$ and a novel transition reward $r_{\textit{trans}}$ which are parameterized by $\theta_{r_p}$ and $\theta_{r_{\textit{trans}}}$. The goal for NRSS is to find the optimal parameters $\theta_{r_p}$ and $\theta_{r_{\textit{trans}}}$ by using the Monte-Carlo tree search thus improving recommendation performance.
However, wireless sensing feedback lacks generalization ability as it is only available for certain tasks or scenarios, making it hard to determine dynamic user interest. 
\paragraph{Value-based methods}
Prior to Q-learning, value iteration is a more traditional value-based reinforcement learning algorithm that focuses on the iteration of the value function. Gradient Value Iteration (GVI)~\cite{zhang2017dynamic} is proposed to improve the traditional value iteration algorithm by utilizing the transition probability and a multi-agent setting to predict chronological author collaborations. It introduces a new parameter named `status' to reflect the amount of knowledge that the agent needs to learn from this state. The policy is updated only when the distance between the new status and the old status is lower than a pre-defined threshold. However, value iteration requires the transition probability, which is hard to obtain in most cases. Hence, Q-learning and its variants are widely used in DRL-based RS. Cascading DQN (CDQN) with a generative user model~\cite{chen2019generative} is proposed to deal with the environment with unknown reward and environment dynamics. The generative user model adopts GANs to generate a user model based on an offline dataset. Different from previous work, it will generate the reward function for each user to explain the users' behavior. The user model can be written as,
\begin{align}
    \argmax_{\phi\in\triangle^{k-1}} \mathbb{E}_{\phi}[r(s_t,a_t)]-R(\phi)/\eta
\end{align}
where $\triangle^{k-1}$ is the probability simplex, $R(\phi)$ is the regularization term for exploration and $\eta$ is a constant.

Pseudo Dyna-Q (PDQ)~\cite{zou2020pseudo} points out that Monte-Carlo tree search may lead to an extremely large action space and an unbounded importance weight of training samples. Hence, a world model is proposed to reduce the instability of convergence and high computation cost for interacting with users by imitating the offline dataset. With the world model, the agent will interact with the learned world model instead of the environment to improve the sample efficiency and convergence stability. The world model learning process introduced in PDQ can be described as finding the parameter $\theta_M$,
\begin{align}
    \argmin_{\theta_M}\mathbb{E}_{\xi\in P_{\xi}^\pi}[\sum_{t}^{T-1}\gamma^t\prod_{j=0}^t\frac{\pi(s_j,a_j)}{\pi_b(s_j,a_j)}\Delta_t(\theta_M)]
\end{align}
where $\xi$ is generated by the logged policy $\pi_b$, $\prod_{j=0}^t\frac{\pi(s_j,a_j)}{\pi_b(s_j,a_j)}$ is the ratio used for importance sampling and $\Delta$ is the difference between the reward in the world model and real reward. 
Furthermore, GoalRec~\cite{wang2021reinforcement} designs a disentangled universal value function to be integrated with the world model to help the agent deal with different recommendation tasks. The universal value function is defined as
\begin{align}
    V_{\pi} (s)= \mathbb{E}[\sum_{t=0}^\infty r(s_t,a_t)\prod_{k=0}^t\gamma s_k|s_0=s].
\end{align}
Moreover, GoalRec introduces a new variable goal $g\in G$ used to represent users' future trajectory and measurement $m\in M$. $m$ is an indicator that reflects users' response to the given future trajectory based on historical behaviors. Based on that, the optimal action will be selected based on
\begin{align}
    a^* = \max_a U(M(s,a),g)
\end{align}
with a customized liner function $U(\cdot)$.
\begin{figure}
    \centering
    \includegraphics[width=\linewidth]{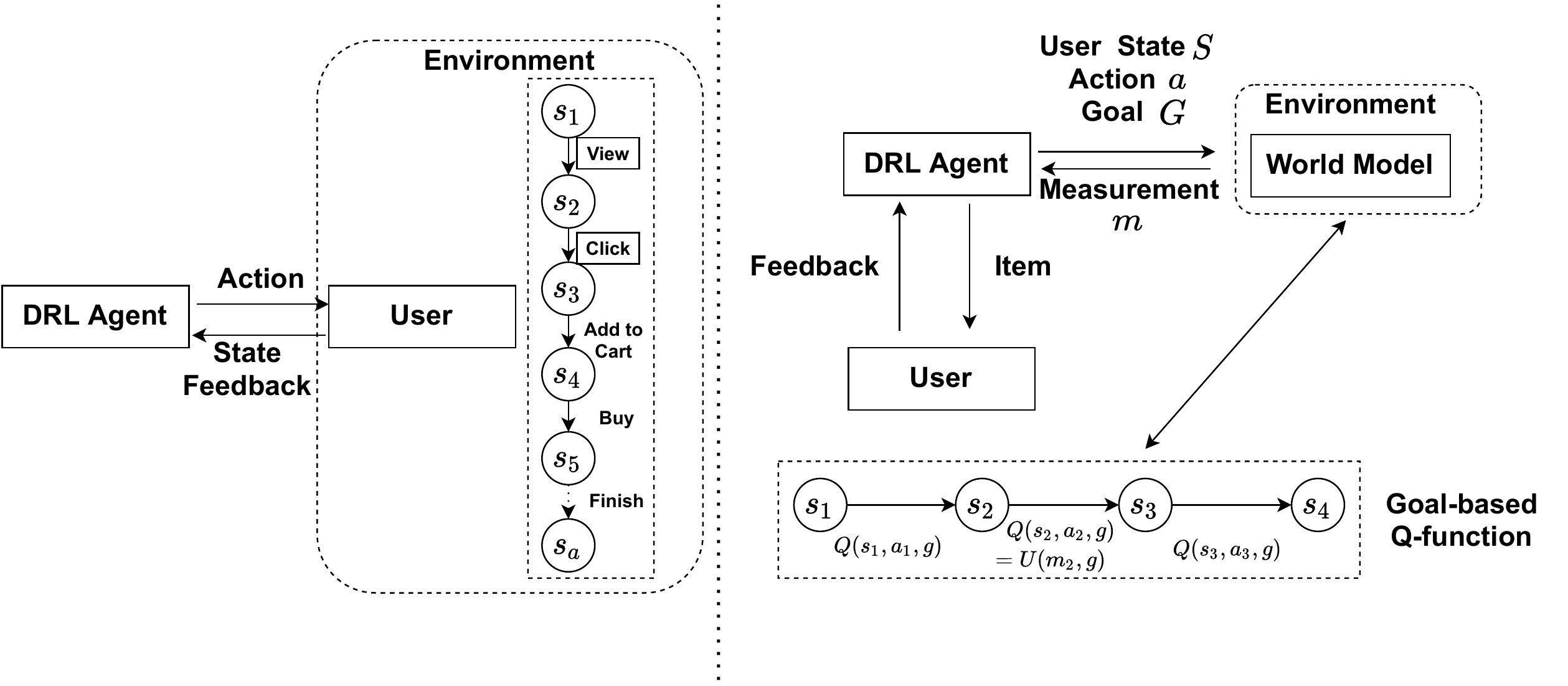}
    \caption{Left is the general structure of model-free methods. Right is the structure for GoalRec which is a model-based method. A sample trajectory is used to demonstrate the difference between them~\cite{wang2021reinforcement}. }
    \label{fig:worldmodel}
\end{figure}

\paragraph{Hybrid methods}
Hybrid method can be recognized as a midpoint between value-based and policy gradient-based methods. DeepChain~\cite{zhao2020whole} uses the multi-agent setting to relieve the sub-optimality problem. The sub-optimality problem is caused by the \textit{one for all} setting that optimizes one policy for all users. Hence, DeepChain designs a multi-agent setting that adopts several agents to learn consecutive scenarios and jointly optimizes multiple recommendation policies. The main training algorithm used is DDPG. To this end, users' actions can be formulated in a model-based form as follows:
\begin{align}
    \sum_{m,d}[p_m^s (s_t,a_t)\gamma Q_{\theta}(s'_{t},\pi_m(s'_{t})) + p_m^c(s_t,a_t)(r_t+\gamma Q_{\theta}(s'_{t},\pi_d(s'_{t}))) + p_m^l (s_t,a_t)r_t]1_m
\end{align}
where $m$ represents the number of actor networks, $c,l,s$ represent the three different scenarios, $1_m$ is used to control the activation of two actors and $(m,d)\in\{(1,2),(2,1)\}$.
\paragraph{Discussion}
%The benefit of using model-based DRL in RS is that the agent can plan ahead by thinking ahead and distilling the results from planning ahead into the policy. 
Model-based methods aim to learn a model or representation to represent the whole environment so that the agent can plan ahead and receive better sample efficiency.
%However, there are few studies that focus on model-based DRL in RS.
The drawback of such a method is that the ground-truth representation of the environment is unavailable in recommendation scenarios as it dynamically changes, leading to a biased representation. Moreover, model-based methods use the transition probability function $\mathcal{P}$ to estimate the optimal policy. As mentioned, the transition probability function is normally equivalent to users' behavior probability which is hard to determine in a recommender system. Hence, existing works~\cite{zhao2020whole,bai2019model,wang2021reinforcement,zou2020pseudo,chen2019generative,zhang2017dynamic} approximate $\mathcal{P}$ using a neural network or embedding it into the world model. \citet{zhao2020whole} design a probability network to estimate $\mathcal{P}$ while \cite{bai2019model,chen2019generative} uses a GAN to generate user behavior where $\mathcal{P}$ is embedded in the latent space. Different from them,~\cite{wang2021reinforcement,zou2020pseudo} relies on the world model to predict users' next behavior and feed it into the policy learning process.

The challenges of model-based DRL are not widely used in RS and can be summarized into the following facets:

\begin{itemize}
\item $\mathcal{P}$ is hard to determine in real-world recommender systems. 

\item If approximation is used to estimate $\mathcal{P}$, the overall model complexity will substantially increase as it requires approximating two different functions $\mathcal{P}$ and the recommendation policy $\pi$ by using a large amount of user behavior data. 

\item World model-based methods require periodic re-training to ensure the model can reflect user interests in time which increases the computation cost. 

\end{itemize}
\begin{figure}
    \centering
    \begin{subfigure}{0.48\linewidth}
        \includegraphics[width=\linewidth]{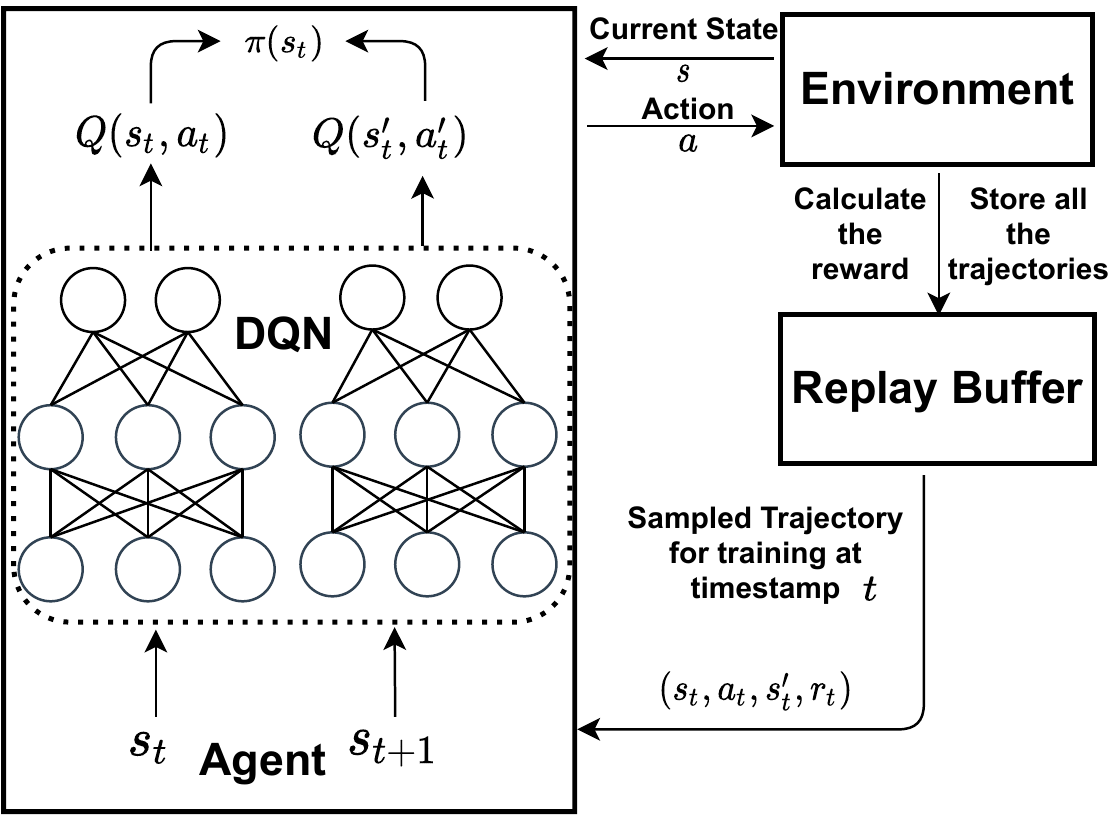}
        \caption{DQN}
    \end{subfigure}
    \begin{subfigure}{0.48\linewidth}
        \includegraphics[width=\linewidth]{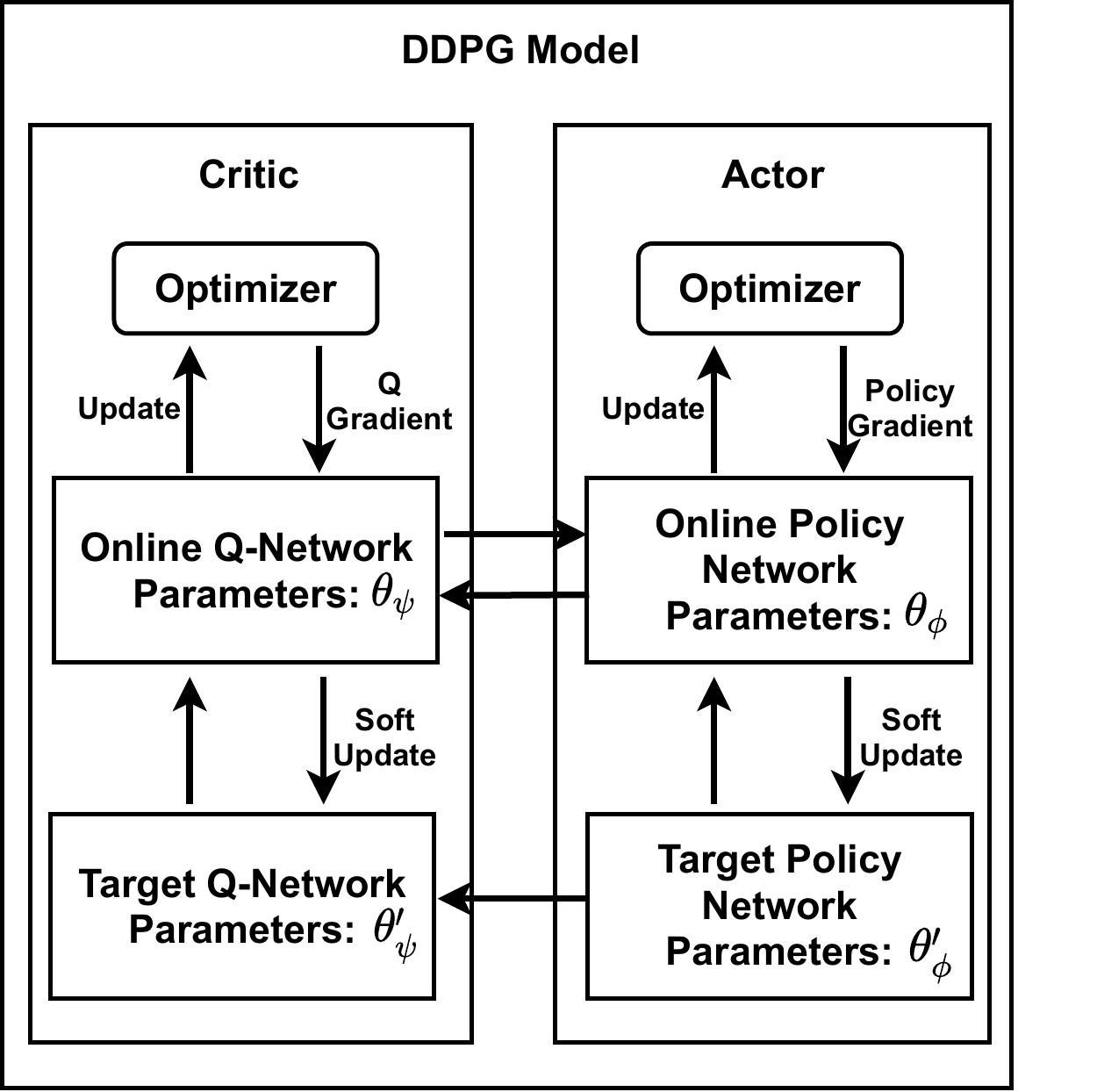}
        \caption{DDPG}
    \end{subfigure}
    \caption{The typical structure of DQN and DDPG}
    \label{fig:ddpg_fig}
\end{figure}
\subsection{Model-free deep reinforcement learning based methods}

Compared with model-based methods, model-free methods are relatively well-studied. Different from model-based methods, $\mathcal{P}$ is unknown and not required in model-free methods. Model-based methods enable the agent to learn from previous experiences. In this subsection, we categorize model-free based DRL in RS into three parts: value-based, policy-based and hybrid methods. 
\begin{table}[ht]
\caption{List of reviewed publications in Model-free DRL-based RS}
\begin{tabular}{ccc}
\hline
Tasks & Note & Work \\\hline
\multirow{4}{*}{Value-based} & Vanilla DQN and its extensions &\cite{zheng2018drn,zhao2018deep,lei2019social} \\ \
& DQN with state/action space optimization & \cite{xiao2020deep,lei2019interactive,zou2019reinforcement,48200} \\
& DQN with graph/image input & \cite{lei2020reinforcement,gui2019mention,oyeleke2018situ,zhao2018recommendations,takanobu2019aggregating,gao2019drcgr,zhou2020interactive} \\
& DQN for joint learning & \cite{pei2019value,zhao2020jointly,zhao2021dear}\\\hline
\multirow{3}{*}{Policy-based} & Vanilla REINFORCE &\cite{pan2019policy,wang2018reinforcement,chen2019top,xu2020reinforcement,ma2020off,chen2021user,montazeralghaem2020reinforcement,ji2020spatio,yu2019vision}  \\
& REINFORCE uses graph structure/input & \cite{wang2020kerl,xian2019reinforcement,wang2020reinforced,chen2019large}\\
& Non-REINFORCE based & \cite{hu2018reinforcement,zhang2019text} \\ \hline
\multirow{2}{*}{Hybrid} & Vanilla DDPG & \cite{zhao2017deep,zhao2018deep,liu2020top,wang2018supervised,cai2018reinforcement} \\
& with Knowledge Graph & \cite{chen2020knowledge,zhao2020leveraging,feng2018learning,zhang2021intelligent,he2020learning,he2020learning,haarnoja2018soft,zhao2020mahrl,xie2021hierarchical}\\

\hline
\end{tabular}
\end{table}
\paragraph{Value based methods}
% DQN, first introduced in RS and its extension
As mentioned, Deep Q-learning and its variants are typical value-based DRL methods widely used in DRL-based RS. DRN~\cite{zheng2018drn} is the first work utilizing Deep Q-Networks (DQN) in RS. It adopts Double DQN (DDQN)~\cite{van2016deep} to build a user profile and designs an activeness score to reflect how frequently a user returns after one recommendation plus users' action (click or not) as the reward. DRN provides a new approach to integrating DRL into RS when dealing with a dynamic environment. The key objective function can be found as follows,
\begin{align}
    \mathbb{E}[r_{t+1} + \gamma Q_{\theta_t'}(s_{t+1}, \argmax_{a'} Q_{\theta_t}(s_t,a'))]
\end{align}
where $a'$ is the action that gives the maximum future reward according to $\theta_t$, $\theta_t$ and $\theta_t'$ are different parameters for two different DQNs. 
\citet{zhao2018deep} points out that negative feedback will also affect recommendation performance which DRN does not consider. Moreover, positive feedback is sparse due to the large number of candidate items in RS. Only using positive feedback would lead to convergence problems. Hence, DEERS is proposed to consider both positive and negative feedback simultaneously by using DQN. Gated Recurrent Units (GRU) are employed to capture users' preferences for both a positive state $s^+$ and negative state $s^-$ and the final objective function can be computed as:
\begin{align}
    \mathbb{E}[r_{t+1} + \gamma \max_{a_{t+1}}Q_{\theta_q}(s^+_{t+1},s^-_{t+1},a_{t+1})|s^+_t,s^-_t,a_t].
\end{align}
\citet{lei2019social} introduces attention mechanisms into the DQN to leverage social influence among users. To be specific, a social impact representation $U_v$ is introduced into the state representation. Matrix factorization is adopted to determine 
%the similar
similarity among
users and hence present the social influence. Social attention is introduced to distill the final state representation.
% How DQN used in state space optimization.
In addition, a few studies focus on user profiling to improve recommendation performance~\cite{xiao2020deep,lei2019interactive,zou2019reinforcement}. \citet{lei2019interactive} claims that user feedback contains useful information in the previous feedback even when the user does not like the recommended items. Some existing studies focus on final feedback which ignore the importance from earlier steps to later ones. Hence, user-specific DQN (UQDN) is proposed to consider multi-step feedback from users. It employs Matrix Factorization to generate user-specific latent state spaces. The newly defined objective function with the user-specific latent state space can be represented as
\begin{align}
    \mathbb{E}[r_{t+1} + \gamma \max_{a_{t+1}}\overline{Q}_{\theta_q}(s_{t+1},a_{t+1}) + \overline{\textbf{b}}_u - Q_{\theta_q} (s_{t+1},a_{t+1})]
\end{align}
where $\overline{\textbf{b}}_u$ is the learned user latent representation.
\citet{zou2019reinforcement} also points out that most studies do not consider users' long-term engagement in the state representation as they focus on the immediate reward. FeedRec is proposed that combines both instant feedback and delayed feedback into the model to represent the long-term reward and optimize the long-term engagement by using DQN. To be specific, time-LSTM is employed to track users' hierarchical behavior over time to represent the delayed feedback which contains three different operations: $h_{\mathit{skip}},h_{\mathit{choose}},h_{\mathit{order}}$. The state space is the concatenation of those operations and users' latent representation.  Differently, \citet{xiao2020deep} focuses on the user privacy issue in recommender systems. Deep user profile perturbation (DUPP) is proposed to add perturbation into the user profile by using DQN during the recommendation process. Specifically, DUPP adds a perturbation vector into users' clicked items as well as the state space, which contains users' previous behavior.

% How DQN used in action space optimization 
Distinct from previous studies which focus on optimizing user profiles or state spaces, some studies aim to optimize the action space formed by interactions with items. In the situation of basket recommendation, the user is suggested multiple items as a bundle, which is called a recommendation slate. It leads to combinatorially large action spaces making it intractable for DQN based recommendation models.

SlateQ~\cite{48200} is proposed to decompose slate Q-value to estimate a long-term value for individual items, and it is represented as,
\begin{align}
    Q_{\theta_q}(s_t,a_t)  = \sum_{i \in a_t}p(i|s_t,a_t)\overline{Q}_{\theta_q}(s_t,i)
\end{align}
where $\overline{Q}_{\theta}(s,i)$ is the decomposed Q-value for item $i$. The decomposed Q-value will be updated by the following rule which is similar to traditional DQN,
\begin{align}
    \overline{Q}_{\theta_q}(s_t,i) \leftarrow \alpha \bigg(r_t +\gamma \sum_{j\in a_{t+1}}p(j|s_{t+1},a_{t+1}) \overline{Q}_{\theta_q}(s_{t+1},j)\bigg) + (1-\alpha)\overline{Q}_{\theta_q}(s_t,i).
\end{align}
Different from other mode-free methods, Slate-Q assumes that the transition probability $p(i|s_t,a_t)$ is known.

% How DQN combined with other techniques to solve the complex data model
Vanilla DQN methods may not have sufficient knowledge to handle complex data such as images and graphs. \citet{tang2018personalized} firstly models users' click behavior as an embedding matrix in the latent space to include the skip behaviors of sequence patterns for sequential recommendation. Based on that, \citet{gao2019drcgr} propose DRCGR, which adopts CNN and GAN into DQN to help the agent to better understand high-dimensional data, e.g., a matrix. Two different convolution kernels are used to capture users' positive feedback. In the meantime, DRCGR uses GANs to learn a negative feedback representation to improve robustness. 
% However, it is not clear why two kernels are required.
Another typical data format is the graph, which is widely used in RS, including knowledge graphs.
\citet{lei2020reinforcement} propose GCQN which adopts Graph Convolutional Networks (GCN)~\cite{kipf2017semi} into the DQN which constructs the state and action space as a graph-aware representation. Differently, GCQN introduces the attention aggregator: $\sum_{w\in \mathcal{N}(i)}\alpha_{iu}e_u$ which demonstrates better performance than the mean-aggregator and pooling-aggregator. For item $i$, the graph-aware representation can be represented as,
\begin{align}
    \sigma\bigg(W_{fc}[e_i\oplus\sum_{w\in \mathcal{N}(i)}\alpha_{iu}e_u + b_{fc}]\bigg)
\end{align}
where $W_{fc},b_{fc}$ are the parameters for the fully-connected layer, $e_u$ is the embedding for user $u$ and $\mathcal{N}(i)$ is the set of one-hot neighbours of item $i$ in graph $G(i)$. \citet{zhou2020interactive} propose KGQR uses a similar strategy to transform the information into a knowledge graph which is fed into the GCN to generate the state representation. Notably, KGQR presents a different state representation generation method. For given node $i$, the neighbourhood representation with a $k$-hop neighborhood aggregator can be represented as,
\begin{align}
    e_i^k = \sigma\bigg(W_{k}\frac{1}{|\mathcal{N}(i)|}\sum_{t\in\mathcal{N}(i)} e_{t}^{k-1} + B_k e_i^{k-1}\bigg)
\end{align}
where $\mathcal{N}(i)$ is the set of neighboring nodes, $W_k, B_k$ are the parameter of the aggregator. Those neighbourhood representations will be fed into a GRU and the state representation will be generated. Another application domain for using graph data is job recommendation which requires considering multiple factors jointly such as salary, job description, job location etc. SRDQN~\cite{sun2021cost} constructs a probability graph to represent a candidate's skill set and  employs a multiple-task DQN structure to process these different factors concurrently. 
% How DQN used when applying RS into e-commerce platform - advertising?

%Recommendation may associate with advertising when deployed into e-commerce platform.
There are some studies targeting recommendation and advertising simultaneously in e-commerce environments~\cite{pei2019value,zhao2020jointly,zhao2021dear}. \citet{pei2019value} mentions when deploying RS into real-world platforms such as e-commerce scenarios, the expectation is to improve the profit of the system.
%which most of existing studies do not cover. 
A new metric, Gross Merchandise Volume (GMV), is proposed to measure the profitability of the RS to provide a new view about evaluating RS in advertising. Different from GMV, \citet{zhao2020jointly} separates recommendation and advertising as two different tasks and proposes the Rec/Ads
Mixed display (RAM) framework. RAM designs two agents: a recommendation agent and an advertising agent, where each agent employs a CDQN to conduct the corresponding task. \citet{zhao2021dear} find that advertising and recommendation may harm each other and formulate a rec/ads trade-off. Their proposed solution, DEARS, contains two RNNs. Two RNNs are employed to capture user preferences toward recommendations and ads separately. Based on that, DQN is employed to take those two outputs as the input to construct the state and output the advertising.

\paragraph{Policy-based methods}
% vanilla PG
Policy-based DRL can be divided into two parts which are Constrained Policy Optimization (CPO) ~\cite{achiam2017constrained} and policy gradient.
%There is only one work that uses CPO in RS.
\citet{zhang2019text} uses CPO to identify the contradiction between text feedback and historical preferences. It provides a solution for using DRL in the situation where users' feedback is entirely different from previous feedback in RS. 
Policy gradient-based methods are the other stream in policy-based DRL methods for RS. These methods aims to optimize the policy $\pi$ directly instead of estimating the Q-value like DQN. A well-known and widely used policy gradient method in RS is REINFORCE which uses the following rule for policy $\pi_{\theta_\pi}$,
\begin{align}
    \theta \leftarrow \theta + \alpha \mathbb{E}_{\tau\sim d_{\pi_{\theta_\pi}}}\bigg[\sum_{t=1}^Tr(s_t^i,a_t^i)\sum_{t=1}^T\nabla_{\theta_\pi} \log \pi_{\theta_\pi}(s_t^i,a_t^i)\bigg]
\end{align}
where $i$ is sampled trajectories from $\pi_{\theta}(a_t|s_t)$. \citet{pan2019policy} propose Policy Gradient for Contextual Recommendation (PGCR), which adopts REINFORCE and considers contextual information. PGCR assumes that the policy follows the multinoulli distribution, in which case the transition probability can be estimated easily through sampling from previously seen context.

\citet{wang2018reinforcement} incorporate CNNs and attention mechanisms in REINFORCE for explainable recommendation. Specifically, this work designs a coupled agent structure where one agent generates the explanation and the other makes recommendations based on the generated explanation.

\citet{chen2019top} increases the scalability of REINFORCE to ensure it can deal with the extremely large action space under recommendation scenarios. To be specific, it introduces a policy correction gradient estimator into REINFORCE to reduce the variance of each gradient by doing importance sampling. The new update rule
becomes
%can be shown as,
\begin{align}
    \theta_\pi \leftarrow \theta_\pi + \alpha \sum_{\tau\sim\beta}\bigg[\sum_{t=1}^T\frac{\pi_{\theta_\pi}(s_t,a_t)}{\pi_\beta(s_t,a_t)}r(s_t^i,a_t^i)\sum_{t=1}^T\nabla_{\theta_\pi} \log \pi_{\theta_\pi}(s_t^i,a_t^i)\bigg]
\end{align}
where $\pi_\beta$ is the behavior policy trained by state-action pairs without the long-term reward and $\pi_{\theta}$ is trained based on the long-term reward only. It is worth mentioning that the vanilla REINFORCE algorithm is on-policy, and importance sampling will make REINFORCE behave like an off-policy method with the following gradient format,
\begin{align}
    \mathbb{E}_{\tau \sim d_{\pi_\theta}}\bigg[\prod_{t'=1}^t\frac{\pi_{\theta}(s_t,a_t)}{\pi_{\theta_s}(s_{t'},a_{t'})}\sum_{t'=t}^Tr(s_t,a_t)\sum_{t=1}^T\nabla_{\theta} \log \pi_{\theta}(s_t,a_t)\bigg]
\end{align}
where $\pi_{\theta_s}$ is the sample policy parameter.
\citet{xu2020reinforcement} also finds that the REINFORCE method suffers from a high variance gradient problem and Pairwise Policy Gradient (PPG) is proposed. Different from policy correction, PPG uses Monte Carlo sampling to sample two different actions $a,b$ and compare the gradient to update $\theta$,
\begin{align}
    \mathbb{E}_{\tau\sim d_{\pi_{\theta_\pi}}}\bigg(\sum_a\sum_b(r(s,a)-r(s,b))\sum_{t=1}^T(\nabla_{\theta_\pi}\log_{\pi_{\theta_\pi}}(s_t,a_t)-\nabla_{\theta_\pi}\log_{\pi_{\theta_\pi}}(s_t,b_t))\bigg).
\end{align}
\citet{ma2020off} extends the policy correction gradient estimator into a two-stage setting which are $p(s_t,a^p)$ and $q(s_t,a|a^p)$ and the policy can be written as
\begin{align}
    \sum_{a^p}p(s_t,a^p)q(s_t,a|a^p).
\end{align}
In addition, weight capping and self-normalized importance sampling are used to further reduce the variance.
Moreover, a large state space and action space will cause  sample inefficiency problems as REINFORCE relies on the current sampled trajectories $\tau$. \citet{chen2021user} finds that the auxiliary loss can help improve the sample efficiency%as it is well-studied in recent literature for relieving such problems
~\cite{jaderberg2016reinforcement,sermanet2018time}. Specifically, a linear projection is applied to the state $s_t$, the output is combined with action $a_t$ to calculate the auxiliary loss and appended into the final overall objective function for optimization.

Another prototype of vanilla policy gradient in DRL-based RS is the policy network. \citet{montazeralghaem2020reinforcement} designs a policy network to extract features and represent the relevant feedback that can help the agent make a decision. Similar to DQN, this work uses a neural network to approximate the Q-value and the policy directly without theoretical analysis.
\citet{ji2020spatio} extend the policy network by introducing spatio-temporal feature fusion to help the agent understand complex features. Specifically, it considers both the current number and the future number of vacant taxis on the route to recommend routes for taxis. \citet{yu2019vision} introduces multi-modal data as new features to conduct vision-language recommendation by using historical data to train REINFORCE. ResNet and attention are used to encode vision and text information, respectively. Moreover, two rewards are introduced with a customized ratio $\lambda$ to balance vision and text information. 
% PG with graph

Knowledge Graphs (KG) are widely used in RS to enrich side information, provide explainability and improve recommendation performance. Similar to DQN, vanilla REINFORCE cannot properly handle graph-like data. \citet{wang2020kerl} propose a method named Knowledge-guided Reinforcement Learning (KERL), which integrates  knowledge graphs into the REINFORCE algorithm. To be specific, KERL adopts TransE~\cite{bordes2013translating} to transfer the knowledge graph into a graph embedding and utilizes a multilayer perceptron (MLP) to predict future knowledge of user preferences. The state representation can be written as
\begin{align}
    h_t \oplus \mathit{TransE}(\mathcal{G}) \oplus \mathit{MLP}(\mathit{TransE}(\mathcal{G}))
\end{align}
where $h_t$ is the hidden representation from the GRU for sequential behavior and $\mathcal{G}$ is the knowledge graph.

Different from KERL, \citet{xian2019reinforcement} propose Policy-Guided Path Reasoning (PGPR), which formulates the whole environment as a knowledge graph. The agent is trained to find the policy to find good items conditioned on the starting user in the KG by using REINFORCE. PGPR uses the tuple $(u,e_t,h_t)$ to represent the state instead of the graph embedding where $e_t$ is the entity the agent has reached at $t$ for user $u$ and $h_t$ is the previous action before $t$.
%With such a tuple-like state representation, the action requires adjustment to fulfill the graph structure. 
The action in PGPR is defined as the prediction of all outgoing edges for $e_t$ based on $h_t$.

\citet{wang2020reinforced} propose a knowledge graph policy network (KGPolicy) which puts the KG into the policy network and adopts REINFORCE to optimize it. In addition, KGPolicy uses negative sampling instead of stochastic sampling to overcome the false negative issue---sampled items behave differently during training and inference. Similar to GCQN, attention is also employed to establish the representation for its neighbors.

Due to the on-policy nature of REINFORCE, it is difficult to apply it to large-scale RS as the convergence speed will be a key issue. To relieve this, \citet{chen2019large} propose TPGR, which designs a tree-structured policy gradient method to handle the large discrete action space hierarchically. TPGR uses balanced hierarchical clustering to construct a clustering tree. Specifically, it splits a large-scale data into several levels and maintains multiple policy networks for each level to conduct the recommendation. The results are integrated at the final stage.

% DPG 
As mentioned, policy gradient can be further extended to deterministic policy gradient (DPG) \cite{silver2014deterministic}. \citet{hu2018reinforcement} propose Deterministic Policy Gradient with Full Backup
Estimation (DPG-FBE) to complete a sub-task of recommendation. DPG-FBE considers a search session MDP (SSMDP) that contains a limited number of samples, where the stochastic policy gradient method like REINFORCE cannot work well.
\paragraph{Hybrid methods}
% DDPG
The most common model-free hybrid method used would be the actor-critic algorithm where the critic network uses the DQN and the actor uses the policy gradient. The common algorithm used to train actor-critic is DDPG with the following objective function,
\begin{align}
    \mathbb{E}[r_t+\gamma Q_{\theta_q'}(s_{t+1},\mu_{\theta_\pi'}(s_{t+1})) - Q_{\theta_q}(s_t,a_t)]
\end{align}
where $\theta_q,\theta_q'$ is the parameter for Q-learning at time $t,t+1$ while $\theta_\pi'$ is the parameter for deterministic policy gradient at time $t+1$.
\citet{zhao2017deep} propose LIRD, which uses the vanilla actor-critic framework to conduct list-wise recommendations. In order to demonstrate the effectiveness of LIRD, a pre-trained user simulator is used to evaluate the effectiveness of LIRD where the transition probability is approximated using the cosine similarity for a given state-action pair $s_t,a_t$.
\citet{zhao2018deep} further extend LIRD into  page-wise recommendation and proposed DeepPage. Similar to other previous work, GRU is employed to process the sequential pattern. Moreover, similar to DRCGR, DeepPage formulates the state as a page, then CNNs are employed to capture features and fed to the critic network. The final state representation is the concatenation of the sequential pattern and the page features. Additionally, there are a few studies focusing on different scenarios such as top-aware recommendation~\cite{liu2020top}, treatment recommendation~\cite{wang2018supervised}, allocating impressions~\cite{cai2018reinforcement} etc. \citet{liu2020top} introduces a supervised learning module (SLC) as the indicator to identify the difference between the current policy and historical preferences. SLC will conduct the ranking process to ensure the recommendation policy will not be affected by the positional bias -- the item appearing on top receives more clicks. Similarly, \citet{wang2018supervised} also integrates the supervised learning paradigm into DRL but in a different way. An expert action $\hat{a}_t$ is provided when the critic evaluates the policy and the update rule is slightly different than normal DQN,
\begin{align}
    \theta_q \leftarrow \theta_q + \alpha \sum_{t}[ Q_{\theta_q}(s_t,\hat{a}_t)-r_t-\gamma Q_{\theta_{q'}}(s_t,\mu_{\theta_{\pi'}}(s_t))]\nabla_{\theta_q}Q_{\theta_q}(s_t,a_t).
\end{align}
However, such a method is not universal as the acquisition of expert action is difficult and depends on the application domain.
%DDPG with KG

Similar to policy gradient and DQN, Knowledge Graphs (KG) are also used in actor-critic-based methods. \citet{chen2020knowledge} propose KGRL to incorporate the substantial information of knowledge graphs to help the critic to better evaluate the generated policy. A knowledge graph is embedded into the critic network. Different from previous studies which use the KG as the environment or state representation, KGRL uses KG as a component in the critic, which can guide the actor to find a better recommendation policy by measuring the proximity from the optimal path. Specifically, a graph convolutional network is used to weight the graph and Dijkstra's algorithm is employed to find the optimal path for finally identifying the corresponding Q-value. \citet{zhao2020leveraging} claim that human's demonstration could improve path searching and propose ADAC. ADAC also searches for the optimal path in the KG but further adopts adversarial imitation learning and uses expert paths to facilitate the search process. \citet{feng2018learning} propose MA-RDPG, which extends the standard actor-critic algorithm to deal with multiple scenarios by utilizing a multi-actor reinforcement learning setting. Specifically, two different actor-networks are initialized while only one critic network will make the final decision. Those two actor networks can communicate with each other to share information and approximate the global state. \citet{zhang2021intelligent} find that there are multiple factors can affect the selection of electric charging station. Hence, it uses a similar idea to recommend the electric vehicle charging station by considering current supply, future supply, and future demand. \citet{he2020learning} figure out that the communication mechanism in MA-RDPG will harm actors as they are dealing with independent modules, and there is no intersection. Hence, \citet{he2020learning} extend MA-RDPG into multi-agent settings which contain multiple pairs of actors and critics and remove the communication mechanism to ensure independence. 
%Moreover, 

Different from \cite{feng2018learning}, \citet{he2020learning} use `soft' actor-critic (SAC)~\cite{haarnoja2018soft}, which introduces a maximum entropy term $\mathcal{H}(\pi(s_t,\phi_t))$ to actor-critic to improve exploration and stability with the stochastic policy $\pi(s_t,\phi_t)$. 
Similar to the multi-agent idea, \citet{zhao2020mahrl} use a hierarchical setting to help the agent learn multiple goals by setting multiple actors and critics. In comparison, hierarchical RL uses multiple actor-critic networks for the same task. It splits a recommendation task into two sub-tasks: discovering long-term behavior and capturing short-term behavior. The final recommendation policy is the combination of the optimal policies for the two sub-tasks. Similarly, \citet{xie2021hierarchical} use the hierarchical setting for integrated recommendation by using different sourced data. The objective is to work out the sub-polices for each source hierarchically and form the final recommendation policy afterward.

\paragraph{Discussion}
In RS, model-free methods are generally more flexible than model-based methods as they do not require knowing the transition probability. We summarize the advantages and disadvantages of the three kinds of methods described under the model-free category. DQN is the first DRL method used in RS, which is suitable for small discrete action spaces. The problems with DQN in RS are: 
\begin{itemize}
    \item RS normally contains large and high-dimensional action spaces.
    \item The reward function is hard to determine which will lead to inaccurate value function approximation.
\end{itemize} 

Specifically, the high dimensional action space in context of recommender systems is recognized as a major drawback of DQN~\cite{mnih2015human,tavakoli2018action}. The reason lies in the large number of the candidate items. Hence, DQN, as one of the most popular schemes, is not the best choice for RS in many situations. Moreover, some unique factors need to be considered when designing the reward function for RS such as social inference. It introduces extra parameters to the Q-network and hinders the convergence.

Policy gradient does not require the reward function to estimate the value function. Instead, it estimates the policy directly. However, policy gradient is designed for continuous action spaces. More importantly, it will introduce high variance in the gradient. Actor-critic algorithms combine the advantages of DQN and policy gradient. Nonetheless, actor-critic will map the large discrete action space into a small continuous action space to ensure it is differentiable, which may cause potential information loss. Actor-critic uses DDPG and thus inherits disadvantages from DQN and DPG, including difficulty in determining the reward function and poor exploration ability.

\subsection{Component Optimization in Deep Reinforcement Learning based RS}
\label{sec:component}
There are a few studies that use DRL in RS for goals other than improving recommendation performance or proposing new application domains. We split the literature based on the following components: environment, state representation, and reward function. Exisitng studies usually focus on optimizing one single component in the DRL setting (as illustrated in~\Cref{fig:overview}).
\begin{table}[h]
    \centering
    \caption{List of publications reviewed in this section}
    \begin{tabular}{c|c}
    \hline
      Component  &  Work \\\hline
       Environment  & \cite{shi2019pyrecgym,rohde2018recogym,shi2019virtual,ie2019recsim,shang2019environment,huang2020keeping,santana2020mars,zhao2019toward} \\
       State & \cite{liu2018deep,liu2020end,liu2020state}\\
       Reward & \cite{chen2018stabilizing}\\
       \hline
    \end{tabular}
    \label{tab:component}
\end{table}
\\
\subsubsection{Environment Simulation and Reconstruction}~\\
Many environments are available for evaluating deep reinforcement learning. Two popular ones are OpenAI gym-based environment \cite{brockman2016openai} and  MuJoCo\footnote{http://mujoco.org/}.
Unfortunately, there is no standardized simulation platform or benchmark specific to reinforcement learning based recommender systems. Existing work on DRL in RS is usually evaluated through offline datasets or via deployment in real applications. The drawback for evaluating offline datasets include:

% Unlike traditional deep learning evaluation processes, there is a lack of standard environments for evaluating DRL in RS.
%The major issue 
%inside 
%is the environment construction. Different from traditional DRL methods for which all methods are evaluated in the OpenAI gym\footnote{https://gym.openai.com/} based environment such as MuJoCo\footnote{http://mujoco.org/}, there is no standardized simulation platform that can be used to evaluate RS. 

\begin{itemize}
    \item Different studies use different environment construction methods which leads to unfair comparison. For instance, some studies use the KG as the environment while some studies assume the environment is gym-like or design a simulator for specific tasks.
    \item With offline datasets, users' dynamic interests, and environment dynamics are hard to maintain. Deploying the method into a real application is difficult for academic research as it takes time and costs money. Hence, a standardized simulation environment is a desirable solution.
\end{itemize}
%comes to a reasonable solution. 
\begin{figure}
    \centering
    \includegraphics[width=\linewidth]{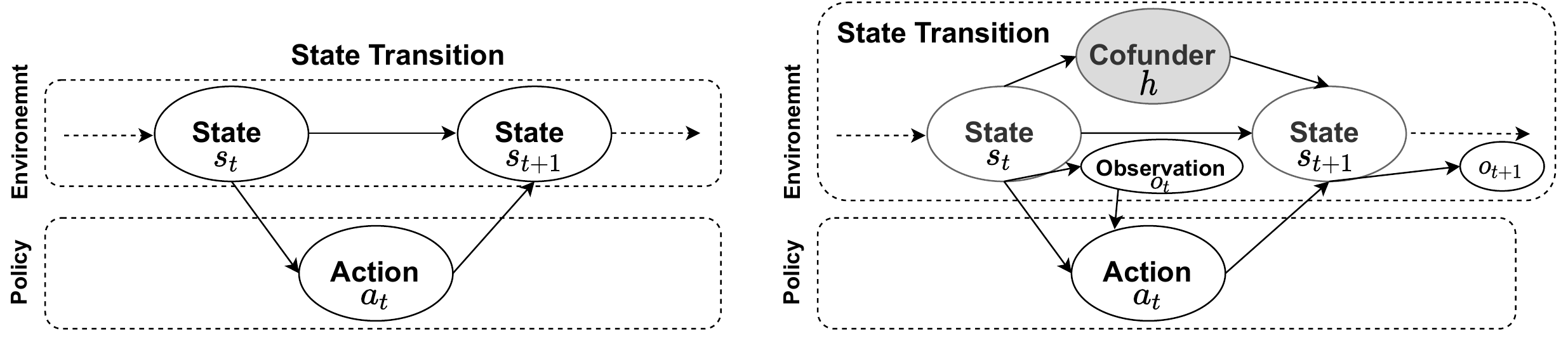}
    \caption{Left is the traditional MDP transition; Right is the POMDP which considers the environmental confounders such as social influence~\cite{shang2019environment}.}
    \label{fig:cofounder}
\end{figure}

There are several different studies that provide standardized gym\footnote{https://gym.openai.com/}-based simulation platforms for DRL-based RS research in different scenarios. RecSim~\cite{ie2019recsim} is a configurable platform that supports sequential interaction between the system and users. RecSim contains three different tasks: interest evolution, interest exploration and long-term satisfaction. RecoGym~\cite{rohde2018recogym} provides an environment for recommendation and advertising. In addition, RecoGym also provides simulation of online experiments such as A/B-tests. However, RecSim and RecoGym are designed for bandit behavior which means users' interests will not change over time. 
%Differently, 
VirtualTB~\cite{shi2019virtual} is proposed to relieve such problems. VirtualTB employs imitation learning to learn a user model to interact with the agent. GANs are employed to generate users' interests. Similar to VirtualTB, Recsimu~\cite{zhao2019toward} uses a GAN to tackle the complex item distribution.
In addition, PyRecGym~\cite{shi2019pyrecgym} accommodates standard benchmark datasets into a gym-based environment. MARS-Gym~\cite{santana2020mars} provides a benchmark framework for marketplace recommendation.~\cite{huang2020keeping} suggests that  existing simulation environments are biased because of  biased logged data. Two common biases are discussed: popularity bias and positivity bias. To reduce the effect from those biases, SOFA introduces an Intermediate Bias Mitigation Step for debiasing purposes.

%Moreover, 
One work discusses environment reconstruction by considering confounders.~\cite{shang2019environment} claims that  users' interests may be affected by social networks which may introduce extra bias to the state and affect the decision-making process. A multi-agent setting is introduced to treat the environment as an agent which can partially relieve the hidden confounder effect. Specifically, a deconfounded environment reconstruction method DEMER is proposed. Different from previously  mentioned methods, DEMER assumes the environment is partially observed and models the whole recommendation task as a Partially Observed MDP (POMDP). Different from an MDP, a POMDP contains one more component observation $o\in\mathcal{O}$ and the action $a_t$ is derived based on the observation $o_t$ instead of the state $s_t$ by $a_t = \pi_a(o_t)$. DEMER assumes there is a confounder policy $\pi_h$ for observation $o_h$ which is composed by $a_t$ and $o_t$: $a_h = \pi_h(a_t,o_t)$. Moreover, another observation $o_b$ is introduced to observe the transition as well. $\pi_b$ is the corresponding policy and $a_b = \pi_b(o_b) = \pi_b(o_t,a_t,a_h)$. DEMER uses generative adversarial imitation learning (GAIL) to imitate the policy $A,B$.
Given trajectory $\{o_t,o_h,o_b\}$ for different policies $A$ and $B$, the objective function is defined as
\begin{align}
    & (\pi_a,\pi_b,\pi_h) = \argmin_{(\pi_a,\pi_b,\pi_h)}\mathbb{E}_{s\sim \tau }(L(s,a_t,a_b)) \notag \\
    & \text{where }L(s,a_t,a_b) = \mathbb{E}_{(\pi_a,\pi_b,\pi_h)}[\log D(s,a_t,a_b)]-\lambda \sum_{\pi\in\{\pi_a,\pi_b,\pi_h\}}H(\pi)
\end{align}
where $L(\cdot)$ is the loss function, $D(\cdot)$ is a discriminator and $H(\pi)$ is introduced in GAIL.
\\
\subsubsection{State Representation}
\begin{figure}[!h]
    \centering
    \includegraphics[width=0.5\linewidth]{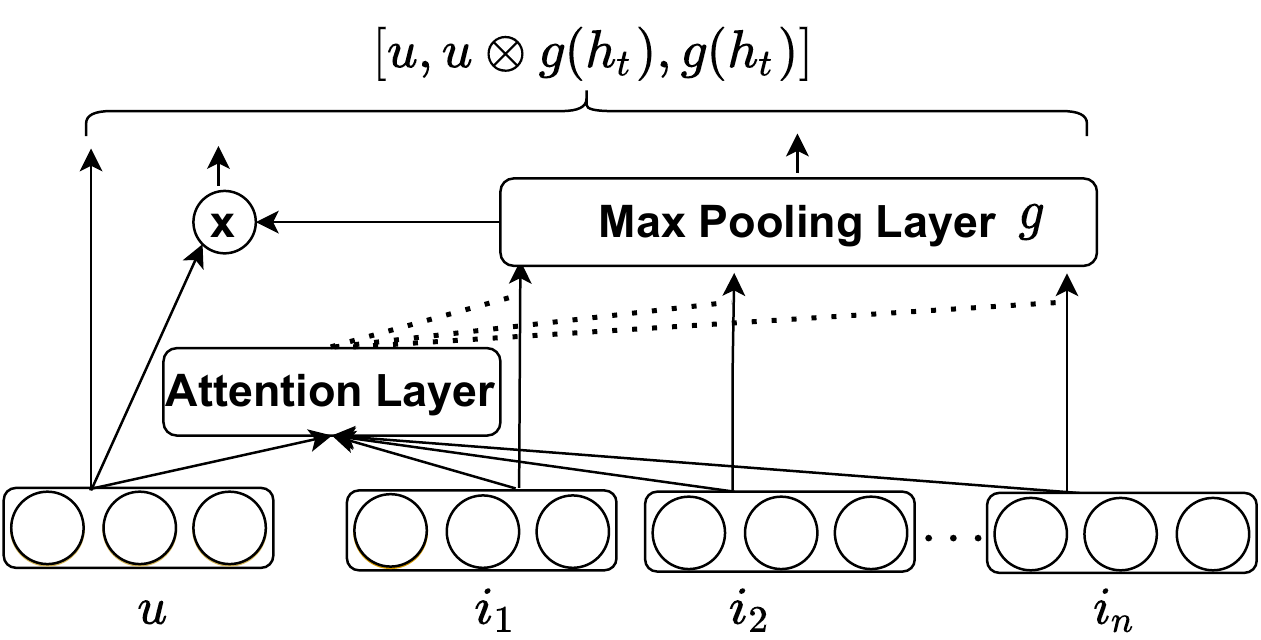}
    \caption{State representation used in works~\cite{liu2020end,liu2020state}. $h_t$ is the output of an attention layer that takes the representation of users' history at time $t$ as input and $g(\cdot)$ is the pooling operation. }
    \label{fig:staterep}
\end{figure}

~\\
State representation is another component in DRL-based RS which exists in both model-based and model-free methods. \citet{liu2018deep} find that the state representation in model-free methods would affect  recommendation performance.
%certainly. 
%Benefiting from the world model, model-based methods are less affected significantly but still some. It is caused by the state construction methods where most of 
Existing studies usually directly use the embedding as the state representation. \citet{liu2020end,liu2020state} propose a supervised learning method to generate a better state representation by utilizing an attention mechanism and a pooling operation as shown in~\Cref{fig:staterep}.
Such a representation method requires training a representation network when training the main policy network, which increases the model complexity.
\\
\subsubsection{Robustness of Reward Functions}~\\
The reward function is crucial for methods involving DQN.
%involved methods.
A robust reward function can significantly improve training efficiency and performance. \citet{kostrikov2018discriminatoractorcritic} find that the DQN may not receive the correct reward value when entering the absorbing state. That is, when the absorbing state is reached, the agent will receive zero reward and harm policy learning. The reason behind this is that when designing the environment zero reward is implicitly assigned to the absorbing state as it is hard to determine the reward value in such a state. \citet{chen2018stabilizing} propose a robust DQN method, which can stabilize the reward value when facing the absorbing state. The new reward formula can improve the robustness, which is defined as follows:
%,
\begin{align}
    r = \begin{cases} 
      r_t & \text{if } s_{t+1} \text{ is an absorbing state} \\
      r_t + \gamma Q_{\theta'}(s_{t+1},a_{t+1})& \text{otherwise}.
   \end{cases}
\end{align}
The major difference is that $r_t$ is assigned to the absorbing state to ensure the agent can continue learning.
%The above reward function can improve the robustness but still have room to improve. 
One remaining problem in current DRL-based RS is the reward sparsity, i.e,. the large state and action spaces make the reward sparsity problem more serious. One of the possible solution would be a better designed reward by using the reward shaping~\cite{ng1999policy}.

\section{Emerging Topics}
While existing studies have established a solid foundation for DRL-based RS research, this section
outlines several promising emerging research directions.
\subsection{Multi-Agent and Hierarchical Deep Reinforcement Learning-based RS}
Recommender systems are monolithic systems containing 
%lots for 
tasks such as searching, ranking, recommendation, advertising, personalization, and diverse stakeholders such as users and items. Most existing methods are based on single agent. 

% This requires RS to have the capacity to handle multiple tasks at the same time.
%While most existing studies deal with recommendation only which is not realistic in the real application as it normally associates with other tasks.
Multi-Agent Reinforcement Learning (MARL) is a subfield of reinforcement learning and it is capable of learning multiple policies and strategies. 

While a single-agent reinforcement learning framework can only handle a single task, many studies consider the multi-task situation in RS and employ multi-agent DRL (MADRL) or hierarchical DRL (HDRL).
%would be a better approach. 
HDRL~\cite{kulkarni2016hierarchical} is proposed to handle complex tasks by splitting such tasks into several small components and asks the agent to determine sub-policies. HDRL belongs to a single-agent reinforcement learning framework such that the agent contains a meta-controller and several controllers. The meta-controller splits the task, and the controllers learn the value and reward functions for designated tasks to get a series of sub-policies. There are a few studies already utilizing HDRL in RS.
 \citet{xie2021hierarchical} target integrated recommendation to capture user preferences on both heterogeneous items and recommendation channels. Specifically, the meta-controller is used for item recommendation, and controllers aim to find the personalized channel according to user channel-level preferences. \citet{zhang2019hierarchical} uses HDRL for course recommendation in MOOCs, which contains two different tasks: profile reviser and recommendation. The meta-controller aims to make course recommendations by using the revised profile pruned by the controllers.

Different from HDRL, MADRL~\cite{egorov2016multi} introduces multiple agents to handle the sub-tasks. \citet{gui2019mention} uses the MADRL for twitter mention recommendation where three agents are initialized. The three agents need to generate different representations for the following tasks: query text, historical text from authors and historical text from candidate users. Once the representations are finalized, the model will conduct the recommendation based on the concatenation of representations. \citet{feng2018learning} and \citet{he2020learning} provide two different views of the communication mechanism in MADRL and demonstrate that agents could work collaboratively or individually. \citet{zhao2020jointly} designs a MADRL framework for two tasks where two agents are designed to conduct advertising and recommendation respectively. \citet{zhang2017dynamic} uses MADRL for collaborative recommendation where each agent is responsible for a single user. MADRL is adopted to help the recommender consider both collaboration and potential competition between users. \citet{zhang2021intelligent} designs a charging recommender system for intelligent electric vehicles by using decentralized agents to handle sub-tasks and a centralized critic to make the final decision.
% Several reviewed works~\cite{shang2019environment} employ the MADRL to conduct two or more tasks when make recommendation and achieve some improvement compare to single-agent based methods. HDRL can decompose a big task into several sub-task, agent needs to work out several sub-policies hierarchically. 
% zhang2021intelligent

Hierarchical multi-agent RL (HMARL)~\cite{makar2001hierarchical} proves that MARL and HRL can be combined.
%and achieves a good performance. 
Recently, \citet{yang2018hierarchical} introduces HMADRL into the continuous action space, which provides a direction for RS. \citet{zhao2020mahrl} uses HMARL for multi-goal recommendation where the meta-controller considers users' long-term preferences and controllers focus on short-term click behavior. While the meta-controller and controllers in HDRL deal with sub-tasks that belong to a single task, HMARL focuses on multi-task or multi-goal learning where the meta-controller and controllers belong to different agents and deal with different tasks or goals.

HMADRL would be a suitable solution for future research work in DRL-based RS where HDRL can be used to split a complex task into several sub-tasks such as users' long-term interests and short-term click behavior, and MADRL can jointly consider multiple factors such as advertising \citet{zhao2020mahrl}.

\subsection{Inverse Deep Reinforcement Learning for RS}
As
%aforementioned, 
mentioned,
the reward function plays a critical role in DRL-based recommender systems. In many existing works, reward functions are manually designed. The common method uses users' click behavior to represent the reward and to reflect users' interests. However, such a setting can not represent users' long-term goals~\cite{zou2019reinforcement} as clicking or not only depicts part of the feedback information from users. It requires significant effort to design a reward function, due to the large number of factors that can affect users' decision, such as social engagement or bad product reviews, which may adversely affect recommendation performance. It is difficult to include all potential factors into the reward function because not every factor  can be represented properly. A few works~\cite{gong2019exact,chen2020generative} show that manually designed reward functions can be omitted by employing inverse reinforcement learning (IRL)~\cite{ng2000algorithms} or generative adversarial imitation learning (GAIL)~\cite{ho2016generative}. Such inverse DRL-based methods require using expert demonstration as the ground truth. However, expert demonstration is often hard to obtain for recommendation scenarios.
%However, expert demonstration does not exist in real-world recommendation.
Those two studies conduct experiments in an offline dataset-based simulation environment that can access expert demonstration. In contrast, \citet{chen2020generative} use IRL as the main algorithm to train the agent while \citet{gong2019exact} use both demonstration and reward to train the agent. \citet{zhao2020leveraging} also employ GAIL to improve recommendation performance. In this work, GAIL is used to learn the reasoning path inside the KG to provide side information to help the agent learn the policy. Although IRL achieves some progress in RS, the lack of demonstration is a key shortcoming that impedes adoption in RS. One possibility is to use the IRL method in casual reasoning to help improve interpretability~\cite{bica2020learning} thus boosting recommendation performance. Alternately, IRL may be suitable for learning users' long-term and static behavior to support the reward function.
%Users' long-term and static behavior, it may not change frequently so that it is possible to get enough trajectories or past demonstrations for IRL to learn.

\subsection{Graph Neural Networks for Boosting DRL-based RS}
Graph data and KG are widely used in RS.
%in recent decade. 
%The majority 
Graph modeling enables an RS to leverage interactions between users and the recommender for reasoning or improving interpretability. According to existing studies about deep learning-based RS~\cite{zhang2019deep}, embedding is a technique used to get the representation for the input data. Graph embedding is a common solution to handle graph-like data. GCN is a type of graph embedding method which
%receives increasing research interest rethe cently. Similarly, GCN and its variants 
are broadly used in RS to process  graph data. \citet{wang2019knowledge} propose a variant of GCN to learn the embedding for KG. Specifically, they propose knowledge graph convolutional networks (KGCN) to capture the high-order structural proximity among entities in a knowledge graph. 

In DRL-based RS, graph data are handled similarly---s underthe transformed into an embedding and fed to the agent. \citet{wang2020kerl} uses a traditional graph embedding method TransE~\cite{bordes2013translating} to generate the state representation for DRL-based RS. There are several studies that use GCN in DRL for recommendations under different settings. \citet{Jiang2020Graph} propose a graph convolutional RL (DGN) method which integrates the GCN into the Q-learning framework for general RL problems by replacing the state encoding layer with the GCN layer. \citet{lei2020reinforcement} extend this method into the deep learning field and apply it to recommender systems. To be specific, multiple GCN layers are employed to process the sub-graphs for a given item $i$. \citet{chen2020knowledge} employs KG inside the actor-critic algorithm to help the agent learn the policy. Specifically, the critic network contains a GCN layer to give weight to the graph and conduct searches in the graph to find an optimal path and hence guide the optimization of policy learning. However, such a method is relatively computationally expensive as it requires jointly training the GCN and the actor-critic network. \citet{gong2019exact} adopts a Graph Attention Network (GAT)~\cite{velivckovic2017graph} into the actor-critic network to conduct  recommendation. In addition, the GAT is used as an encoder to obtain a state representation.

%It is not hard to find that 
A common way of using GCN or its variants in DRL-based RS is the state encoder. %Based on this, there 
The related challenge is the difficulty for the environment to provide a graph-like input to the GCN.
%which means customized environments are expected. 
%In DRL research, the environment should be standardized so that the input provided should be usable for any DRL-based method. We believe such a problem would be crucial for future studies. DGN and its extensions would be a possible direction as it can avoid the modification to the environment.

\subsection{Self-Supervised DRL-based RS}
Self-supervised learning (SSL) is a technique in which the model is trained by itself without external label information. SSL-DRL is receiving growing interest in robotics 
%which demonstrates a outstanding performance
~\cite{kahn2018self,zeng2018learning}. \citet{kahn2018self} shows that SSL can be used to learn the policy when doing navigation by providing real-world experience. \citet{zeng2018learning} demonstrates that SSL-DRL can be used to help the agent learn synergies between two similar policies, thus empowering the agent to conduct two different tasks. 
Recent advances in SSL RL show that SSL can also provide interpretability for RL, which is promising for interpretable RS research~\cite{shi2020self}. \citet{shi2020self} shows that SSL based RL can highlight the task-relevant information to guide the agent's behavior.
Moreover, \citet{xin2020self} shows that SSL can be used to provide negative feedback for DRL-based RS to improve recommendation performance. To be specific, a self-supervised loss function is appended into the normal DRL loss function,
\begin{align}
    -\sum_{i=1}^nY_i\log\bigg(\frac{e^{y_i}}{\sum_{i'=1}^ne^{y_{i'}}}\bigg) + L_{\mathit{DRL}}
\end{align}
where $Y_i$ is an indicator function to show users interact with the item $i$ or not. $L_{\mathit{DRL}}$ could vary, if the DQN is adopted, \Cref{dqnloss} should be used.
SSL demonstrates promising performance in visual representation in recent years, which would be a possible solution to generate the state representation as there are a few DRL-based RS studies that adopt CNNs to process image-like data and transform it into a state~\cite{liu2020top,gao2019drcgr}. Furthermore, as an unsupervised learning approach, SSL would provide a new direction about defining the reward function by learning common patterns between different states as well as multi-task learning.

\section{Open Questions}
In this section, we outline several open questions and challenges that exist in DRL-based RS research. We believe these issues could be critical for the future development of DRL-based RS.
\subsection{Sample Efficiency}
%As mentioned, most existing studies are model-free.
Sample inefficiency is a well-known challenge in model-free DRL methods. Model-free DRL requires a significant number of samples as there is no guarantee that the received state is useful. Normally, after a substantial number of episodes, the agent may start learning as the agent finally receives a useful state and reward signal. A common solution is the experience replay technique, which only works in off-policy methods. 
%Hence, off-policy methods with experience replay are more popular than on-policy methods in DRL and DRL-based RS research. 
%However, 
Experience replay still suffers the sample inefficiency problem~\cite{schaul2015prioritized} as not every past experience is worth replaying. \citet{isele2018selective} propose selected experience replay (SER) that only stores valuable experience into the replay buffer and thus improves sample efficiency. while traditional DRL environments only contain several\footnote{For example, the number of actions in MuJoCo is less than one hounded.} candidate items, in DRL-based RS, the agent must deal with a significantly larger action space as RS may contain lots of candidate items. Existing DRL-based RS studies on traditional experience replay methods often demonstrate slow converge speed. \citet{chen2021user} design a user model to improve the sample efficiency through auxiliary learning. Specifically, they apply the auxiliary loss with the state representation, and the model distinguishes low-activity users and asks the agent to update the recommendation policy based on high-activity users more frequently.
%which could be the possible direction.
%Moreover, 
%We find that those advance in DRL research about experience method can provides some promising direction for improving sample inefficiency problem in model-free based DRL-based RS.

On the other hand, model-based methods are more sample efficient. However, they introduce extra complexity as the agent is required to learn the environment model as well as the policy. Due to the extremely large action space and possibly large state space (depending on users' contextual information) in RS, approximating the environment model and policy simultaneously
%comes to more challenging. 
becomes challenging.
\subsection{Exploration and Exploitation}
The exploration and exploitation dilemma is a fundamental and challenging problem in reinforcement learning research and receives lots of attention in DRL.
%research as well. 
This dilemma describes a trade-off between obtaining new knowledge and the need to use that knowledge to improve performance. Many DQN-based methods focus on exploration before the replay buffer is full and exploitation afterward. Consequently, it requires an extremely large replay buffer to allow all possibilities in recommendation can be stored.
%However, this is not realistic to enumerate all the possibilities in recommendation due to its dynamic nature.}
DRN employs Dueling Bandit Gradient Descent (DBGD)~\cite{yue2009interactively} to encourage exploration while ~\cite{chen2019generative,he2020learning} introduces a  regularization or entropy term into the objective function to do so.~\cite{48200} uses the sheer size of the action space to ensure sufficient exploration.~\cite{wang2020kerl,xian2019reinforcement,wang2020reinforced} uses a separate KG or elaborated graph exploration operation to conduct  exploration.~\cite{chen2019top} employs Boltzmann exploration to get the benefit of exploratory data without negatively impacting user experience. 
In addition, $\epsilon$-greedy is the most common technique used to encourage  exploration ~\cite{zou2020pseudo,lei2020reinforcement,lei2019interactive,zou2019reinforcement,lei2019social,wang2021reinforcement,cai2018reinforcement,liu2020top,xie2021hierarchical}. Remaining studies rely on a simulator to conduct exploration. However, it may suffer from noise and over-fitting~\cite{xie2021hierarchical} because of the gap between simulation and real online application. For most DRL-based methods such as vanilla DQN, policy gradient, or actor-critic-based methods, $\epsilon$-greedy would be a good choice for exploration. In addition, injecting noise into the action space would also be helpful for those actor-critic-based methods~\cite{lillicrap2015continuous}. For methods involving KGs, $\epsilon$-greedy may help, but the elaborated graph exploration methods may receive better performance. 

%%% a bit confusing, just dropped

\subsection{Generalizing from Simulation to Real-World Recommendation}
%\textcolor{red}{
Existing work generally trains DRL algorithms in simulation environments or offline datasets.
Deploying DRL algorithms into real applications is challenging due to the gap between simulation and real-world applications.
% gap
Simulation environments do not contain domain knowledge or social impact.
They can not cover the domain knowledge and task-specific engineering in the real-world recommendation.
%t can be recognized as the gap between the simulation and real-world application.
How to bridge the gap between simulation and real applications is a challenging topic. Sim2real~\cite{zhao2020sim} is a transfer learning approach that transfers DRL policies from simulation environments to reality. Sim2real uses domain adaption techniques to help agents transfer learned policy. Specifically, it adopts GANs to conduct adaption by generating different samples.
RL-CycleGAN~\cite{rao2020rl} is a sim2real method for vision-based tasks. It uses CycleGAN~\cite{zhu2017unpaired} to conduct pixel-level domain adaption. Specifically, it maintains cycle consistency during GAN training and encourages the adapted image to retain certain attributes of the input image.
In DRL-based RS, sim2real would be a possible solution for generalizing the learned policy from simulation environments to reality. However, sim2real is a new technique still under exploration. It shows an adequate capability in simple tasks and requires more effort to handle the complex task such as recommendation. We believe it is a workable solution for generalizing from simulation to reality.
%}
% % resource
% Another reason is the resource shortage.
% %ue to the nature and limitations of model-based methods, model-free methods are more popular in literature. 
% Model-free methods require a significant amount of data from the environment. Benefiting from the simulation platform, the agent can get enough trajectories to support  policy learning. However, it is not possible to train such an agent in real-world application as it is unrealistic to get the same amount of data from users. Moreover, the deployment of DRL algorithms in real applications requires support from industry which is not pervasive. % for some researchers?

\subsection{Bias (Unfairness)}
%With massive data and powerful machine learning models, recommender systems (RS) can achieve strong performance by fitting user behavior data better. However, 
\citet{2010.03240} observe that user behavior data are not experimental but observational, which leads to problems of bias and unfairness.\\
%Bias is so common in RS for the following two reasons.
There are two reasons why bias is so common. First, the inherent characteristic of user behavior data is not experimental but observational. In other words, data that are fed into recommender systems are subject to selection bias \cite{1602.05352}. For instance, users in a video recommendation system  tend to watch, rate, and comment on those movies that they are interested in. Second, a distribution discrepancy exists, which means the distributions of users and items in the recommender system are not even. Recommender systems may suffer from 'popularity bias', where popular items are recommended far more frequently than the others. However, the ignored products in the “long tail” can be equally critical for businesses as they are the ones less likely to be discovered.

% For example, when an online advertisement system places ads, it shows ads that it believes to be of interest to the user but will less frequently display other ads. So population bias arises, and therefore these ads receive more attention from users, and more user behavior is generated in turn. These create a widely recognized problem of bias for recommender systems.\\
\citet{friedman1996bias} denote the unfairness as that the system systematically and unfairly discriminates against certain individuals or groups of individuals in favor of others. 
% it is hard to make recommender systems more entrenched within our society due to the unfairness issue. 
% When training on such unbalanced data, the models are highly likely to learn these over-groups, reinforce them in the ranked results, and potentially result in systematic discrimination.\\

A large number of studies explore dynamic recommendation systems by utilizing the agent mechanism in reinforcement learning (RL), considering the information seeking and decision-making as sequential interactions. How to evaluate a policy efficiently is a big challenge for RL-based recommenders. Online A/B tests are not only expensive and time-consuming but also sometimes hurt the user experience. Off-policy evaluation is an alternative strategy that historical user behavior data are used to evaluate the policy. However, user behavior data are biased, as mentioned before, which causes a gap between the policy of RL-based RS and the optimal policy.

To eliminate the effects of bias and unfairness, \citet{chen2019top} use the inverse of the probability of historical policy to weight the policy gradients. \citet{huang2020keeping} introduce a debiasing step that corrects the biases presented in the logged data before it is used to simulate user behavior. \citet{zou2020pseudo} propose to build a customer simulator that is designed to simulate the environment and handle the selection bias of logged data.

\subsection{Explainability}
Although deep learning-based models can generally improve the performance of recommender systems, they are not easily interpretable. As a result, it becomes an important task to make recommender results explainable, along with providing high-quality recommendations.
%but also convincing explanations.
High explainability in recommender systems not only helps end-users understand the items recommended but also enables system designers to check the internal mechanisms of recommender systems. \citet{INR-066} review different information sources and various types of models that can facilitate explainable recommendation. Attention mechanisms and knowledge graph techniques currently play an important role in realizing explainability in RS.

Attention models have great advantages in both enhancing predictive performance and having greater explainability~\cite{10.1145/3285029}. \citet{wang2018reinforcement} introduce a reinforcement learning framework incorporated with an attention model for explainable recommendation. Firstly, it achieves model-agnosticism by separating the recommendation model from the explanation generator. Secondly, the agents that are instantiated by attention-based neural networks can generate sentence-level explanations.

Knowledge graphs contain rich information about users and items, which can help to generate intuitive and more tailored explanations for the recommendation system \cite{INR-066}. Recent work has achieved greater explainability by using reinforcement and knowledge graph reasoning. The algorithm from \cite{xian2019reinforcement} learns to find a path that navigates from users to items of interest by interacting with the knowledge graph environment. \citet{zhao2020leveraging} extract imperfect path demonstrations with minimum labeling effort and propose an adversarial actor-critic model for  demonstration-guided path-finding. Moreover, it achieves better recommendation accuracy and explainability by reinforcement learning and knowledge graph reasoning.

\subsection{Robustness on Adversarial Samples and Attacks}
Adversarial samples demonstrate that deep learning-based methods are vulnerable. Hence,
robustness becomes an open question for both RS and DRL. Specifically, adversarial attack and defense in RS have received a lot of attention in recent years~\cite{deldjoo2021survey} as security is crucial in RS. Moreover, DRL policies are vulnerable to adversarial perturbations to agent's observations~\cite{lin2017tactics}. \citet{Gleave2020Adversarial} provide an adversarial attack method for perturbing the observations, thus affecting the learned policy. Hence, improving the robustness is the common interest for DRL and RS, which would be a critical problem for DRL-based RS. \citet{cao2020adversarial} provide an adversarial attack detection method for DRL-based RS which uses the GRU to encode the action space into a low-dimensional space and design decoders to detect the potential attack. However, it only considers Fast Gradient Sign Method (FGSM)-based attacks and strategically-timed attacks~\cite{lin2017tactics}. Thus, it lacks the capability to detect other types of attack. Moreover, it only provides the detection method while the defence is still an opening question.
%Attackers can disrupt recommendation systems silently, making defending adversarial attacks a non-trivial task by introducing input noise during interactions.

% The goal of adversarial attacks against DRL-based RS would be to focus on the observations and thus affect the action which follows a certain paradigm. 

We believe zero-shot learning techniques would be a good direction for training a universal adversarial attack detector. For defence, it is still an open question for DRL-based RS, though recent advances in adversarial defence in DRL may provide some insights~\cite{lutjens2020certified,wang2020defense,chen2019adversarial}.

\section{Future Directions}
In this section, we provide a few potential future directions of DRL-based RS. Benefiting from recent advances in DRL research, we believe those topics can boost the progress of DRL-based RS research.
\subsection{Causal and Counterfactual Inference}
Causality is a generic relationship between a cause and effect. Moreover, inferring causal effects is a fundamental problem in many applications like computational advertising, search engines, and recommender systems~\cite{bottou2013counterfactual}.

Recently, some researchers have connected reinforcement learning with learning causality to improve the effects for solving sequential decision-making problems.
Besides, Learning agents in reinforcement learning frameworks face a more complicated environment where a large number of heterogeneous data are integrated.
From our point of view, causal relationships would be capable of improving the recommendation results by introducing the directionality of cause and the effect. The users' previous choices have impact on the subsequent actions. This can be cast as an interventional data generating the dynamics of recommender systems. By viewing a policy in RL as an intervention, we can detect unobserved confounders in RL and choose a policy on the expected reward to better estimate the causal effect \cite{shang2019environment}. Some studies improve RL models with causal knowledge as side information. Another line of work uses causal inference methods to achieve unbiased reward prediction~\cite{10.1145/3397269}.  
% The authors 21`te{shang2019environment} meet a problem of environment reconstruction in reinforcement learning for recommender systems. By taking hidden confounders into consideration that hidden confounders can influence both actions and rewards, they combine RL with causal inference and propose a deconfounded multi-agent environment reconstruction method, and apply it into driver program recommendation system successfully. They make the first attempt in RL to reconstruct the environment together with hidden confounders.

\citet{yang2021causal} propose a Causal Inference Q-network which introduces observational inference into DRL by applying extra noise and uncertain inventions to improve resilience. Specifically, in this work, noise and uncertainty are added into the state space during the training state, and the agent is required to learn a causal inference model by considering the perturbation.
\citet{dasgupta2019causal} give the first demonstration that model-free reinforcement learning can be used for causal reasoning. They explore meta-reinforcement learning to solve the problem of causal reasoning. The agents trained by a recurrent network able to make causal inferences from observational data and output counterfactual predictions. 
%into. 
\citet{forney2017counterfactual} bridge RL and causality by data-fusion for reinforcement learners. Specifically, online agents combine observations, experiments and counterfactual data to learn about the environment, even if unobserved confounders exist. Similarly, \citet{gasse2021causal} make the model-based RL agents work in a causal way to explore the environment under the Partially-Observable Markov Decision Process (POMDP) setting. They consider interventional data and observational data jointly and interprete model-based reinforcement learning as a causal inference problem. In this way, they bridge the gap between RL and causality by relating common concepts in RL and causality.  

Regarding explainability in RL, \citet{madumal2020explainable} propose to explain the behavior of agents in reinforcement learning with the help of causal science. The authors encode causal relationships and learn a structural causal model in RL, which is used to generate explanations based on counterfactual analysis. With counterfactual exploration, this work is able to generate two contrastive explanations for `why' and `why not' questions.

It is so important to search for a Directed Acyclic Graph (DAG) in causal discovery. Considering traditional methods rely on local heuristics and predefined score functions, \citet{zhu2019causal} propose to use reinforcement learning to search DAG for causal discovery. They use observational data as an input, RL agents as a search strategy and output the causal graph generated from an encoder-decoder NN model.
%Casual reasoning guide RL~\cite{cohen2020rational}

\subsection{Offline DRL and Meta DRL}

Recommender systems often need to deal with multiple scenarios such as joint recommendation and adverting, offline DRL and meta DRL provide a promising direction for achieving multiple scenarios at the same time.

Offline DRL is a new paradigm of DRL that can be combined with existing methods such as self-supervised learning and transfer learning to move toward real-world settings. 
Offline DRL~\cite{levine2020offline} (also known as batch DRL) is designed for tasks which contain huge amounts of data.
%which is called the ``data-driven'' paradigm of DRL. 
%The basic idea can be described as, 
Given a large dataset that contains past interactions, offline DRL uses the dataset for training across many epochs but does not interact with the environment.
%which provide the scalability for real-world decision making process such as RS.
%Different from traditional DRL, 
Offline DRL provides a solution that can be generalized to new scenarios as it was trained by a large sized dataset.
Such generalization ability is critical to RSs, which may need to deal with multiple scenarios or multiple customers.
While offline DRL could provide a new direction for DRL-based RS, it still faces a few problems regarding handling the distributional shifts between existing datasets and real-world interactions. 

Meta DRL~\cite{wang2016learning} is defined as 
%to do
meta learning in the filed of DRL. Meta DRL is another approach to help agents to generalize to new tasks or environments. Different from offline DRL, meta DRL contains a memory unit which is formed by the recurrent neural network to memorize the common knowledge for different tasks. Different from offline DRL, meta DRL does not require a large amount of data to train.

\subsection{Further Developments in Actor-Critic Methods}
%Actor-critic methods in RS still have room to improvement. 
An actor-critic method uses the traditional policy gradient method, which suffers from the high variance problem due to the gap between behavior policy (i.e., the policy that is being used by an agent for action select) and target policy (i.e., the policy that an agent is trying to learn).
%as well.
A method commonly used to 
%relief 
relieve
the high variance problem is Advantage Actor-critic (A2C). 
Different from traditional actor-critic methods, A2C uses an advantage function to replace the Q-function inside the critic network. The advantage function $A(s_t)$ is defined as the expected value of the TD-error. %To be specific, 
The new objective function for policy gradient can be written as,
\begin{align}
     \mathbb{E}_{\tau \sim d_{\pi_\theta}}[\sum_{t=1}^T\underbrace{(Q(s_t,a_t) - V(s_t))}_{A(s_t)}\sum_{t=1}^T\nabla_{\theta} \log \pi_{\theta}(s_t,a_t)].
\end{align}
However, A2C still uses DDPG as the main training algorithm, which may suffer function approximation errors when estimating the Q value.
Twin-Delayed DDPG (TD3)~\cite{fujimoto2018addressing} is designed to improve the function approximation problem in DDPG which uses clipped double Q-learning to update the critic. The gradient update can be expressed as,
\begin{align}
     \mathbb{E}_{\tau \sim d_{\pi_\theta}}[\sum_{t=1}^Tr(s_t,a_t) + \gamma \min(Q_1(s_t,a_t+\epsilon), Q_2(s_t,a_t+\epsilon))\sum_{t=1}^T\nabla_{\theta} \log \pi_{\theta}(s_t,a_t)].
\end{align}
where $\epsilon\sim\textit{clip}(\mathcal{N}(0,\sigma,-c,c))$, $\sigma$ is the standard deviation and $c$ is a constant for clipping.  

Another two ways to improve actor-critic methods are Trust Region Policy Optimization (TRPO)~\cite{schulman2015trust} and Proximal Policy Optimization (PPO)~\cite{schulman2017proximal}, which focus on modification of the advantage function. TRPO aims to limit the step size for each gradient to ensure it will not change too much. The core idea is to add a constraint to the advantage function,
\begin{align}
    \frac{\pi(a|s)}{\pi_{old}(a|s)}A(s),
\end{align}
where the KL divergence will be used to measure the distance between the current policy and the old policy is small enough. PPO has the same goal as TRPO which is to try to find the biggest possible improvement step on a policy using the current data. PPO is a simplified version of TRPO which introduces the clip operation,
\begin{align}
    \min\bigg(\frac{\pi(a|s)}{\pi_{old}(a|s)}A(s),\text{clip}
    \bigg(\frac{\pi(a|s)}{\pi_{old}(a|s)}A(s), 1-\epsilon, 1+\epsilon\bigg)A(s)\bigg).
\end{align}

Soft Actor-Critic (SAC)~\cite{haarnoja2018soft} is another promising variant of the actor-critic algorithm and is widely used in DRL research. SAC uses the entropy term to encourage the agent to explore, which could be a possible direction to solve the exploration and exploitation dilemma. Moreover, SAC assigns an equal probability to actions that are equally attractive to the agent to capture those near-optimal policies.
An example of related work \cite{he2020learning} uses SAC to improve the stability of the training process in RS.
%However, there is still space for using SAC in RS. For example, 
%Those mentioned methods are widely used in traditional DRL research but have not received much attention in DRL-based RS. 
%It is understandable that DRL RL is as emerging topic which still under exploration. 
%We believe those advances in DRL would provide an algorithmic insight for future research when designing models for DRL-based RS.

\section{Conclusion}
%Recommender systems are a cross-domain research area that contains significant intersection with different domains and thus produces a few emerging topics. 
In this survey, we provide a comprehensive overview the use of deep reinforcement learning in recommender systems. We introduce a classification scheme for existing studies and discuss them by category. We also provide an overview of such existing emerging topics and point out a few promising directions. 
%of partial of the open questions. 
We hope this survey can provide a systematic understanding of the key concepts in DRL-based RS and valuable insights for future research. 
\bibliographystyle{ACM-Reference-Format}
\bibliography{sample-base}


\end{document}